\documentclass[12pt]{article}
\usepackage{amsmath,amsthm,amssymb,euscript,array,cancel,color}
\usepackage[colorlinks=true]{hyperref}
\usepackage{url}
\usepackage{bbm}

\usepackage[paper=a4paper,dvips,top=2.54cm,left=2.34cm,right=2.34cm,
foot=1cm,bottom=2.54cm]{geometry}

\newcounter{multieqs}

\newcommand{\be}{\begin{equation}}
	\newcommand{\ee}{\end{equation}}
\newcommand{\eq}[1]{(\ref{#1})}

\newcommand{\bra}[1]{\langle #1|}
\newcommand{\ket}[1]{|#1 \rangle}
\newcommand{\ipr}[2]{\langle #1 | #2 \rangle}

\def\one{\mbox{1 \kern-.59em {\rm l}}}

\def\g{\gamma}      
\def\d{\delta}      
\def\e{\epsilon}          
\def\f{\phi}          
\def\h{\eta}  
  
\def\l{{\lambda}}   
\def\m{\mu} \def\n{\nu}  
\def\o{\omega}  
\def\p{\pi}   
\def\r{\rho}  
\def\s{\sigma}    
\def\t{\tau}

\def\z{\zeta}

 \def\cH{{\cal H}}   
  \def\cL{{\cal L}}

\def\cV{{\cal V}}

\def\Ht{\tilde{H}}

\def\Jt{\tilde{J}}

\def\d{\delta}  

\def\pa{\partial}

\def\lrar{\leftrightarrow}

\def\hlf{\frac{1}{2}}  
\def\ove#1{\frac{1}{#1}}

\def\bcomment#1{}  
  
\def\VEV#1{\left\langle #1\right\rangle}

\newcommand{\lrbrk}[1]{\left(#1\right)}

\def\w{{\wedge}}

\def\paP{\pa P}
\def\paPt{\widetilde{\pa P}}

\numberwithin{equation}{section}

\author{Sujiphat Janaun
	\footnote{sujiphatj62@nu.ac.th}$~^{a,b}$\,,
	\,
	Anajak Phonchantuek\footnote{anajakp61@nu.ac.th}$~^a$\,\\
	and Pichet Vanichchapongjaroen
	\footnote{pichetv@nu.ac.th}$~^a$
	\\
	\\
	{\small $^a$ \it The Institute for Fundamental Study ``The Tah Poe Academia Institute",}
	\\
	{\small\it Naresuan University, Phitsanulok 65000, Thailand}
	\\
	\\
	{\small $^b$ \it Department of Physics, Faculty of Science,}
	\\
	{\small\it Naresuan University, Phitsanulok 65000, Thailand}
}

\title{Nonlinear chiral forms in the Sen formulation}

\begin{document}
\maketitle

\abstract{The Sen formulation for chiral $(2p)$-form in $4p+2$ dimensions describes a system with two separate sectors, one is physical while the other is unphysical. Each contains a chiral form and a metric. In this paper, we focus on the cases where the self-duality condition for the unphysical sector is linear while for the physical sector can be nonlinear. We show the decoupling at the Hamiltonian and Lagrangian levels. The decoupling at the Hamiltonian level follows the idea in the literature. Then by an appropriate field redefinition of the corresponding first-order Lagrangian, the separation at the Lagrangian level follows. We derive the diffeomorphism of the theory in the case where the chiral form in the physical sector has nonlinear self-dual field strength and couples to external $(2p+1)$-form field. Explicit forms of Sen theories are also discussed. The Lagrangian for the quadratic theory is separated into two Henneaux-Teitelboim Lagrangians. We also discuss the method of generating explicit nonlinear theories with $p=1$. Finally, we also show that the M5-brane action in the Sen formulation is separated into a Henneaux-Teitelboim action in unphysical sector and a gauge-fixed PST in physical sector.}

\section{Introduction}
Chiral $(2p)$-form fields in $4p+2$ dimensional spacetime with $p=0,1,2,\cdots$ have been extensively studied in the literature partly due to its importance in string theory and M-theory. By definition, each of these fields has a self-dual $(2p+1)$-form field strength.

The self-duality condition makes it non-trivial to construct the action for the corresponding chiral $(2p)$-form field. Several approaches are developed in the literature. For a review of key developments, see for example section 2.1 of \cite{Andriolo:2023ltv}.

In this work, we are particularly interested in the Sen formulation of chiral forms \cite{Sen:2015nph}, \cite{Sen:2019qit} which is inspired from the Sen formalism for string field theory \cite{Sen:2015uaa}. The original form of this formulation can be explained as follows. An action describing chiral $(2p)$-forms is constructed out of a $(2p)$-form field $P$ and a $(2p+1)$-form field $Q$.
Even in the curved spacetime with metric $g$, the field $Q$ is self-dual with respect to a flat metric $\h$. The theory can also contain other external fields collectively called $\Psi$.
With the analysis at the Hamiltonian and the equation of motion level \cite{Sen:2015nph}, \cite{Sen:2019qit}, \cite{Andriolo:2020ykk}, it is found that the Sen theory describes two chiral $(2p)$-form fields, one is physical while the other is unphysical due to the incorrect sign of the kinetic term. The physical chiral field is chiral with respect to the standard metric $g$. The unphysical chiral field is chiral with respect to $\h$ and is decoupled from the physical chiral field and the standard metric $g$.

Although this work only focuses on the classical aspects, it is worth mentioning that it is important to justify that the Sen theory also works at the quantum level. Important works in this directions are carried out in \cite{Sen:2015nph}, \cite{Andriolo:2021gen}.

Let us now come back to the classical aspect. A recent development \cite{Hull:2023dgp} suggests that the flat metric can be promoted to another metric $\bar{g}.$ There are two frameworks on this promotion.
One of the frameworks is that the metric $\bar{g}$ is still taken to be a fixed background metric,
while the other framework is that the metric $\bar{g}$ can be promoted to become dynamical.
In this paper, we will only focus on this second framework.
As a consequence, the Sen theory is separated into two sectors. One of the sectors is physical. It contains a physical chiral $(2p)$-form field, a standard metric $g$, and other physical external fields $\Psi$. The other sector is unphysical. It contains an unphysical chiral $(2p)$-form field and the unphysical metric $\bar{g}$. The analysis at the equation of motion level suggests that these two sectors are decoupled from each other.

The Lagrangian level of the Sen formulation is still not completely understood.
An important difficulty comes from the highly nontrivial way that the
fields $P$ and $Q$ couple to the metric $g$ at the Lagrangian level. This difficulty is present even in the case of linear self-dual theories.
Although some insights has been gained
from the construction of the explicit form of the action \cite{Sen:2015nph}, \cite{Sen:2019qit}, \cite{Andriolo:2020ykk}, \cite{Hull:2023dgp}, \cite{Vanichchapongjaroen:2020wza},
the difficulty still largely remains.

Another difficulty is that it seems not clear how the separation of the two sectors can be realised at the Lagrangian level, in contrast to the well-understood separation at the Hamiltonian and equation of motion levels. A hint that the separation at the Lagrangian level is possible comes from the study of dimensional reduction of Sen theories
\cite{Sen:2015nph}, \cite{Sen:2019qit}, \cite{Andriolo:2020ykk}, \cite{Phonchantuek:2023iao}.
For example, it is found that
under double dimensional reduction,
the Lagrangian of M5-brane in Sen formulation
\cite{Vanichchapongjaroen:2020wza} is reduced to be the Lagrangian which is separated into two sectors.
The unphysical sector describes the free Maxwell gauge field which is uncoupled from all of the physical fields.
The physical sector can either be described by the standard D4-brane action or the dual D4-brane action.
This is to be compared with the double dimensional reduction of the M5-brane action in the PST formulation \cite{Pasti:1995ii}, \cite{Pasti:1996va}, \cite{Pasti:1996vs}, \cite{Pasti:1997gx}, \cite{Bandos:1997ui}
giving rise to the dual D4-brane action.

Therefore, it is to be expected that the separation of the Lagrangian level of the Sen formulation can be achieved. 
It is also hoped that this would lead to the understanding of how the Sen formulation and the PST formulation are related at the Lagrangian level.
The relationship itself is quite non-trivial due to the interchange of the roles of on-shell and off-shell conditions.
In the PST formulation, the $(2p+1)$-form self-dual field is exact off-shell and is self-dual on-shell.
On the contrary, in the Sen formulation the $(2p+1)$-form self-dual field is self-dual off-shell and is exact on-shell.

In this paper, the main aim is to better understand the Lagrangian level of Sen theories in the case where the self-duality in the physical sector can be nonlinear. In particular, we will show that the separation between the physical and unphysical sectors can also be made at the Lagrangian level, in addition to the separation at the Hamiltonian and equation of motion levels obtained in the literature.
As to be discussed in this paper, it turns out that the separation at the Lagrangian level indeed provides a further insight. For example, the relationship between the Sen formulation and the gauge-fixed PST formulation \cite{Henneaux:1988gg}, \cite{Perry:1996mk}, \cite{Aganagic:1997zk}, \cite{Aganagic:1997zq} at the Lagrangian level can be seen explicitly.

This paper is organised as follows. In section \ref{sec:review}, we review important features of the Sen formulation and focus mostly on the Lagrangian level. In section \ref{sec:conventions}, we introduce useful conventions to be used in the analysis in this paper.
Although our main interest is at the Lagrangian level,
it turns out that much insight is gained from the study at the Hamiltonian level. Technical complications arise from the complicated way that the fields $P$ and $Q$ are coupled to the standard metric $g$. The presence of another metric $\bar{g}$ also adds to the complication.
The conventions are introduced to make the calculation at the Hamiltonian level more manageable.
In section \ref{sec:analysis}, we study properties of a general form of Sen theory. For example, in subsection \ref{subsec:Hamgen},
we investigate the condition to be imposed on the Lagrangian so that the Hamiltonian is separated into the two sectors.
In subsection \ref{subsec:dualdecoup}, we work out the separation of the two sectors at the Lagrangian level.
In section \ref{sec:quad}, we work out explicitly how the two sectors of quadratic Sen theory with general metrics $g, \bar{g}$ are separated at the Hamiltonian and Lagrangian levels.
In section \ref{sec:nl6d}, we consider explicit forms of nonlinear action in the Sen formulation in the case $p = 1.$ We will describe how to generate all the nonlinear chiral $2$-form theories in $6$ dimensional spacetime.
We will also show explicitly how the M5-brane action in the Sen formulation is separated at the Lagrangian level.
In section \ref{sec:conclusion}, we conclude, discuss, and suggest future works.

\section{Sen formulation for chiral fields}\label{sec:review}
In this section, we review 
the Sen formulation for chiral $(2p)$-form in $4p+2$ dimensional spacetime \cite{Sen:2015nph}, \cite{Sen:2019qit}, \cite{Andriolo:2020ykk}, \cite{Hull:2023dgp}.

One important key feature of this formulation is that
each Sen theory describes a system which is separated into two sectors uncoupled from each other. One sector is called the physical sector. It contains the standard metric $g$, a chiral $(2p)$-form field which is chiral with respect to $g$. This sector may also contain other physical external fields $\Psi$. 
The other sector is called the unphysical sector. It contains the unphysical metric $\bar{g}$ and an unphysical chiral $(2p)$-form field which is chiral with respect to $\bar{g}.$

Each Sen theory describes dynamics of a
$(2p)$-form $P$ and a $(2p+1)$-form $Q$ which satisfies $Q = \bar{*}Q,$ where $\bar{*}$ is the Hodge star with respect to an external unphysical metric $\bar{g}.$ The Lagrangian of a general Sen theory is of the form
\be\label{Sen-gen}
\begin{split}
\cL = \frac{(2p)!}{2}\bigg(&\ove{4}dP\w\bar{*} dP - Q\w dP \\
&\ +\frac{2}{(2p)!}\cL_I(Q,g,\bar{g},\Psi)d^{4p+2}x\bigg),
\end{split}
\ee
where
\be\label{dQLI}
\d_Q\cL_I(Q,g,\bar{g},\Psi)
=\frac{(2p)!}{2}\d Q\w R(Q,g,\bar{g},\Psi),
\ee
and we have suppressed $d^{4p+2}x$ on LHS of eq.\eq{dQLI}.
This suppression will be adopted throughout this paper.
The field $R$ is a $(2p+1)$-form which is anti-self-dual with respect to $\bar{*},$ i.e. $R = -\bar{*}R.$
By considering the equation of motion level \cite{Sen:2015nph}, \cite{Sen:2019qit}, \cite{Andriolo:2020ykk}, \cite{Hull:2023dgp}, it can be seen that the combination
\be
H = Q - R
\ee
describes degrees of freedom of physical chiral $(2p)$-form field. The composite field $H$ should satisfy a nonlinear self-duality condition which should only involve $H$ and physical external fields $g, \Psi.$
By using the self-duality condition, one can completely determine $R.$

It can be seen from the Lagrangian that $P$ partly provides
unphysical degrees of freedom due to the wrong sign of the kinetic term. At the equation of motion level, part of $P$ can be combined with $Q$ to form a field
\be\label{Hs-defn}
H_{(s)} = Q + \ove{2}dP + \ove{2}\bar{*}dP,
\ee
which is closed: $dH_{(s)} = 0.$
The field $H_{(s)}$ is the field strength
of the unphysical chiral $(2p)$-form field.
It is linear self-dual with respect to $\bar{g}.$
The generalisation to nonlinear self-duality in the unphysical sector is possible \cite{Hull:2023dgp}.
However, since we will mostly focus on the physical sector in this work, we leave the unphysical sector with linear self-duality.

Note that\footnote{We thank anonymous reviewers for the important question leading to the discussion in this paragraph and the next four paragraphs.}
we are mainly interested in the 
system of the fields $P, Q$
and treat $g, \bar{g},$ as well as $\Psi$ as external fields.	
However, we keep in mind the framework in which 
$g$ and $\bar{g}$ can both be promoted to be dynamical.
This is one of the frameworks suggested by \cite{Hull:2023dgp}. The other framework is to allow $g$ to be promoted to be dynamical while
always treating $\bar{g}$ as a fixed background.

It is important to discuss the validity of the framework we are following.
In particular,
there is an argument against $\bar{g}$ being promoted to be dynamical.
The field equations of $P$ and $Q$ for 
the Lagrangian \eq{Sen-gen} are given by
\be\label{QP}
d\lrbrk{Q + \hlf\bar{*}dP} = 0,
\ee
\be\label{RP}
R = \hlf(dP - \bar{*}dP).
\ee
By substituting eq.\eq{RP} into eq.\eq{QP},
one obtains
\be
d(Q - R) = 0.
\ee
Since $H = Q-R$ is also non-linear self-dual with respect to $*$,
one then interprets $H$ as the field strength of physical chiral $(2p)$-form field.
The field $H$ itself depends on $Q, g, \bar{g}, \Psi.$
In particular, the dependence on $\bar{g}$ suggests that
$\bar{g}$ which couples to unphysical chiral field
also couples to physical chiral field.
So if $\bar{g}$ is dynamical,
then the physical chiral field would couple to unphysical chiral field,
which is problematic.

We do not agree with this argument.
The process of the substitution of eq.\eq{RP} into eq.\eq{QP}
suggests that $P$ is eliminated and is given in terms of $Q$.
However, this could potentially give incorrect information on the degrees of freedom.
In the equations \eq{QP}-\eq{RP},
only $P$ is differentiated twice with respect to time.
On the other hand $Q$ is differentiated at most once with respect to time.
The initial value problem then suggests that while $P$ and $\dot P$
can be freely determined at the initial time, the initial data of $Q$ and $\dot Q$ is constrained. It is therefore better to eliminate $Q$
as opposed to the elimination of $P$ as what essentially carried out in the previous paragraph.
In subsection \ref{Qelim}, we discuss the consequence of the elimination of $Q$ at the level of the equation of motion. An important upshot is that
the physical chiral field is only coupled to $g$ and external physical field $\Psi$
while the unphysical chiral field is only coupled to $\bar{g}.$ This suggests the complete separation of the two sectors, and hence even if $\bar{g}$ become dynamical it would not couple to the physical sector.

In order to make it completely clear that $\bar{g}$ can become dynamical without issue, one should consider the full system whose Lagrangian also contains the kinetic terms of $g$ and $\bar{g}$ to see if $\bar{g}$ also decouples from the physical chiral field. Since the dependence of the Sen Lagrangian on $g$ and $\bar{g}$ is complicated, the analysis is anticipated to be quite involved. We expect to attempt on this in a future work.

For definiteness, we will work in this paper within
the framework that $\bar{g}$ can be promoted to be dynamical.
We will assume that this framework is valid.
Even if it ultimately turns out that this is not valid,
the results of this paper will not be affected.
This is because the other framework, in which $\bar{g}$ is fixed, is always valid and is less restrictive.
For example, there is no problem with the coupling between $\bar{g}$ and the physical chiral field since $\bar{g}$ has no dynamics.

The role of physical and unphysical sectors can be interchanged \cite{Hull:2023dgp} simply by introducing an overall minus sign to the Lagrangian. We will not consider this interchange in this paper.

In \cite{Sen:2019qit}, it has been shown that by using Hamiltonian analysis of a general Sen theory with $\bar{g} = \h$ that the Hamiltonian of the theory can be separated into the sum of the Hamiltonian of the two sectors.
However, as we will show in subsection \ref{subsec:Hamgen},
when $\bar{g}$ is general a more specific form of $\cL_I$ is required so that the separation at the Hamiltonian level is realised.

In the original Sen theory as well as various generalisations \cite{Andriolo:2020ykk}, \cite{Vanichchapongjaroen:2020wza}, \cite{Lambert:2019diy}, the field $H_{(s)}$ is singlet $\d H_{(s)} = 0$ under several kinds of transformation such as diffeomorphism, supersymmetry, kappa symmetry.

The reference \cite{Hull:2023dgp} provides an insight on this feature at least when concerning diffeomorphism.
Consider the case of general $\bar{g}$. There are in fact two types of diffeomorphism transformations\footnote{At the Lagrangian level, these transformation rules are in fact more complicated than presented here. They also involve additional terms \cite{Andriolo:2020ykk}, \cite{Hull:2023dgp} which vanish upon imposing equation of motion.}.
The first type is called $\z$-transformation.
Physical fields transform as $\d^{\textrm{zeta}}_\z H = \pounds_\z H, \d^{\textrm{zeta}}_\z g = \pounds_\z g, \d^{\textrm{zeta}}_\z \Psi = \pounds_\z\Psi,$
where $\pounds_\z$ is the Lie derivative along vector field $\z.$
On the other hand, unphysical fields --- $H_{(s)}$ and $\bar{g}$ --- are singlet under this transformation. This then makes $P$ and $Q$ transform in a non-standard way.
The second type is called $\chi$-transformation.
Physical fields are singlet under this transformation while unphysical fields transform as $\d^{\textrm{chi}}_\chi H_{(s)} = \pounds_\chi H_{(s)}, \d^{\textrm{chi}}_\chi \bar{g} = \pounds_\chi \bar{g}.$ The standard diffeomorphism is the diagonal subgroup with $\zeta = \chi.$ That is $\d^{\textrm{standard}}_\xi = \d^{\textrm{zeta}}_\xi + \d^{\textrm{chi}}_\xi = \pounds_\xi$ when applied to any field, including $P$ and $Q.$

In the case $\bar{g} = \h$, the symmetry under $\chi$-transformation and hence under standard diffeomorphism transformation is broken. Only the symmetry under $\z$-transformation remains, giving rise to non-standard transformation rule of $P, Q$ such that $H_{(s)}$ is singlet.

Consider a particular example theory in which $H$ is linear self-dual with respect to the standard curved metric $g$ and with the presence of the external $(2p+1)$-form source $J$.
The linear self-duality condition reads
$H^J = *H^J,$ where $H^J = Q - R + J,$
with $*$ being the Hodge star with respect to the standard metric $g.$ The Lagrangian for this theory is
\be\label{cL}
\begin{split}
\cL = \frac{(2p)!}{2}\bigg(&\ove{4}dP\w\bar{*} dP - Q\w dP\\
&\ + \hlf (Q+J)\w R + \hlf Q\w J\bigg),
\end{split}
\ee
where
\be
\d_Q\lrbrk{(Q+J)\w R + Q\w J} = 2\d Q\w R.
\ee
This Lagrangian is invariant under both $\z$- and $\chi$-transformation \cite{Andriolo:2020ykk}, \cite{Hull:2023dgp}. The invariance under gauge transformation
on $P$ is straightforward,
whereas the invariance under gauge transformation on $J$ is not realised unless the gauge transformation parameter is restricted \cite{Andriolo:2020ykk} or appropriate $(4p+2)$-forms are added into the Lagrangian \cite{Hull:2023dgp}, \cite{Vanichchapongjaroen:2020wza}.

Properties of more general theories, in which $H$ is nonlinear self-dual is explored in section \ref{sec:analysis}.

\section{Conventions}\label{sec:conventions}

\subsection{Index notation}

Let Greek indices for example $\m,\n,\r,\cdots$ represent $(4p+2)$-dimensional spacetime indices. So they run in $0,1,2,\cdots, 4p+1.$ The spacetime coordinates are denoted $x^\m.$ We will also denote $t = x^0.$ Let Roman indices for example $a,b,c,i,j,k,\cdots$ represent spatial indices $1,2,\cdots, 4p+1.$
So the index notation for physical and unphysical metrics are $g_{\m\n}$ and $\bar{g}_{\m\n},$ respectively.
We denote $g^{\m\n}$ to be 
the matrix inverse of $g_{\m\n}$
as usual.
Furthermore, we denote 
$\bar{g}^{\m\n}$ to be matrix inverse of $\bar{g}_{\m\n}.$
Note that generically, $\bar{g}^{\m\n} \neq g^{\m\r}\bar{g}_{\r\s}g^{\s\n}.$ Since the theory involves two metrics, raising and lowering of the indices will be carefully specified when needed.

Note that we preserve the notations $g$ and $\bar{g}$ to refer to the metric tensors. The determinants of the matrix forms of these tensors will be denoted $\det(g)$ and $\det(\bar{g})$.

In order to translate differential form to index notation,
the followings are defined for definiteness.
A differential $q-$form is expressed as
\be
\o_{(q)} = \ove{q!}dx^{\m_1}\w\cdots\w dx^{\m_q}\o_{\m_q\cdots\m_1}.
\ee
Interior products and exterior derivatives act from the right.
The wedge product of all coordinate basis 1-form is
\be
dx^{0}\w dx^{1}\w\cdots \w dx^{4p+1} = d^{4p+2}x.
\ee
There are two Hodge star operators $*, \bar{*}$ corresponding to the two metrics $g, \bar{g}.$
In this paper, we only require the expression for the Hodge star operators acting on $(2p+1)$-forms.
They are defined as
\be
\begin{split}
*(dx^{\m_1}\w\cdots \w dx^{\m_{2p+1}})
&=\frac{dx^{\n_1}\w\cdots \w dx^{\n_{2p+1}}}{(2p+1)!\sqrt{-\det(g)}}\\
&\quad\times g_{\n_1\r_1}\cdots g_{\n_{2p+1}\r_{2p+1}}\\
&\quad\times\e^{\r_1\cdots\r_{2p+1}\m_1\cdots\m_{2p+1}},
\end{split}
\ee
\be
\begin{split}
\bar{*}(dx^{\m_1}\w\cdots \w dx^{\m_{2p+1}})
&=\frac{dx^{\n_1}\w\cdots \w dx^{\n_{2p+1}}}{(2p+1)!\sqrt{-\det(\bar{g})}}\\
&\quad\times\bar{g}_{\n_1\r_1}\cdots \bar{g}_{\n_{2p+1}\r_{2p+1}}\\
&\quad\times\e^{\r_1\cdots\r_{2p+1}\m_1\cdots\m_{2p+1}},
\end{split}
\ee
where $\e^{\m_1\cdots\m_{4p+2}}$ is Levi-Civita symbol with $\e^{01\cdots(4p+1)} = 1.$

It is convenient to define
\be
(\paP)_{\m_1\cdots\m_{2p+1}}
=(2p+1)\pa_{[\m_1}P_{\m_2\cdots\m_{2p+1}]},
\ee
where the antisymmetriser is defined with standard weight, for example, $A_{[ab]} = (A_{ab} - A_{ba})/2.$
We also denote
\be
\paPt^{a_1\cdots a_{2p}}
=-\frac{1}{(2p)!}\pa_{[i}P_{b_1\cdots b_{2p}]}\e^{a_1\cdots a_{2p}ib_1\cdots b_{2p}},
\ee
where $\e^{i_1\cdots i_{4p+1}} = \e^{0i_1\cdots i_{4p+1}}$ is Levi-Civita symbol for spatial part. 

In order to work on Hamiltonian analysis,
it is convenient to make use of Arnowitt-Deser-Misner (ADM) decomposition of both metrics. For the decomposition of standard metric,
we consider hypersurfaces of constant $t.$ The $1$-form describing each hypersurface is normalised, using the lapse function $N,$ as $n = -Ndt$ such that $g^{\m\n}n_\m n_\n = -1.$
The embedding of the hypersurface to the spacetime induces	
the mapping which maps the metric $g$ to its pullback
$\g$ on the hypersurface. The projection of $\pa_t$ with respect to $g$
to the tangent of the hypersurface is given by the shift vector with components $N^i.$
With these descriptions, we have
\be
\g_{ij} = g_{ij},\qquad
N^2 = -\ove{g^{00}},\qquad
N^i = -\frac{g^{0i}}{g^{00}},
\ee
which provide the following decomposition
\be\label{ADM_g}
\begin{split}
g_{00} &= -N^2 + \g_{ij}N^i N^j,\qquad
g_{0i} = g_{i0} = \g_{ij}N^j,\\
g_{ij} &= \g_{ij},
\end{split}
\ee
\be
\begin{split}
g^{00} &= -\ove{N^2},\qquad
g^{0i} = g^{i0} = \frac{N^i}{N^2},\\
g^{ij} &= \g^{ij} - \frac{N^i N^j}{N^2},
\end{split}
\ee
and 
\be\label{ADM_sqrtg}
\sqrt{-\det{(g)}} = N\sqrt{\det(\g)},
\ee
where $\g^{ij}$ is matrix inverse of $\g_{ij}.$
See for example \cite{Gourgoulhon:2007ue} for a review on ADM decomposition and details on geometrical interpretation.
As for the decomposition of unphysical metric
$\bar{g}$,
we also consider the hypersurfaces of constant $t$
but the definitions of the induced metric $\bar{\g}$, lapse function $\bar{N}$,
and shift vector $\bar{N}^i$ are now given with respect to $\bar{g}$.
That is
	\be
	\bar{\g}_{ij} = \bar{g}_{ij},\qquad
	\bar{N}^2 = -\ove{\bar{g}^{00}},\qquad
	\bar{N}^i = -\frac{\bar{g}^{0i}}{\bar{g}^{00}},
	\ee
	which provide the following decomposition
\be
\begin{split}
	\bar{g}_{00} &= -\bar{N}^2 + \bar{\g}_{ij}\bar{N}^i \bar{N}^j,\qquad
	\bar{g}_{0i} = \bar{g}_{i0} = \bar{\g}_{ij}\bar{N}^j,\\
	\bar{g}_{ij} &= \bar{\g}_{ij},
\end{split}
\ee
\be
\begin{split}
	\bar{g}^{00} &= -\ove{\bar{N}^2},\qquad
	\bar{g}^{0i} = \bar{g}^{i0} = \frac{\bar{N}^i}{\bar{N}^2},\\
	\bar{g}^{ij} &= \bar{\g}^{ij} - \frac{\bar{N}^i \bar{N}^j}{\bar{N}^2},
\end{split}
\ee
and 
\be
\sqrt{-\det{(\bar{g})}} = \bar{N}\sqrt{\det(\bar{\g})},
\ee
where $\bar{\g}^{ij}$ is matrix inverse of $\bar{\g}_{ij}.$

\subsection{Multi-index notation and bra-ket notation}\label{subsec:mindex}
In the analysis, it will be convenient to make use of multi-index notation and Dirac bra-ket notation.

Let an index with angled bracket $\VEV{\cdot}$ stands for the collection of $2p$ indices. For example $\VEV{a} = (a_1\cdots a_{2p}).$ Let an index with square bracket $[\cdot]$ stands for the collection of $2p+1$ indices.
For example $[i] = (i_1\cdots i_{2p+1}),$ $[\m] = (\m_1\cdots\m_{2p+1}).$ Furthermore,
\be
dx^{\VEV{a}}\equiv dx^{a_1}\w\cdots\w dx^{a_{2p}},
\ee
\be
dx^{[i]}\equiv dx^{i_1}\w\cdots\w dx^{i_{2p+1}}.
\ee
Using multi-index notation, we have
\be\label{sbar-dT}
\begin{split}
\bar{*}(dt\w dx^{\VEV{a}})
=&-\ove{(2p)!}\e_{0\VEV{b}[i]}w^{0\VEV{a},0\VEV{b}} dx^{[i]}\\
&\qquad+ \ove{(2p)!}\e_{0\VEV{b}[i]}w^{0\VEV{a},[i]} dt\w dx^{\VEV{b}},
\end{split}
\ee
\be\label{sbar-dX}
\begin{split}
\bar{*}dx^{[i]}
=&-\ove{(2p)!}\e_{0\VEV{b}[j]}w^{0\VEV{b},[i]} dx^{[j]}\\
&\qquad+ \ove{(2p)!}\e_{0\VEV{b}[j]}w^{[i],[j]}dt\w dx^{\VEV{b}},
\end{split}
\ee
where
\be\label{wMN}
w^{\m_1\cdots\m_{2p+1},\n_1\cdots\n_{2p+1}}
\equiv\sqrt{-\det(\bar{g})}\bar{g}^{[\m_1|\n_1|}\cdots\bar{g}^{\m_{2p+1}]\n_{2p+1}},
\ee
where indices $\m_1\cdots\m_{2p+1}$ are totally antisymmetrised and indices $\n_1\cdots\n_{2p+1}$ are also totally antisymmetrised.
So
\be
w^{\m_1\cdots\m_{2p+1},\n_1\cdots\n_{2p+1}}
=w^{\n_1\cdots\n_{2p+1},\m_1\cdots\m_{2p+1}}.
\ee
We also have
\be\label{paPtaspaP}
\paPt^{\VEV{a}}
= -\ove{(2p+1)!}\e^{0\VEV{a}[i]}(\paP)_{[i]}.
\ee
Let us define
\be
\hat{\bar{\e}}_{\m_1\cdots\m_{4p+2}} \equiv \sqrt{-\det(\bar{g})}\e_{\m_1\cdots\m_{4p+2}},
\ee
which is the Levi-Civita tensor with respect to $\bar{g}.$
Indices of $\hat{\bar{\e}}$ are raised and lowered by the unphysical metric $\bar{g}.$
In the calculations, we will often need the identity
\be\label{preiden-we}
\hat{\bar\e}^{[\m]}{}_{[\r]}\hat{\bar\e}_{[\m]}{}^{[\n]}
=(2p+1)\hat{\bar\e}^{0\VEV{a}}{}_{[\r]}\hat{\bar\e}_{0\VEV{a}}{}^{[\n]}
+\hat{\bar\e}^{[i]}{}_{[\r]}\hat{\bar\e}_{[i]}{}^{[\n]}.
\ee

Let us now make use of Dirac bra-ket notation.
We will suppress the indices of the form $\VEV{a}$ (which might also include index $0$ when applicable).
Quantities with one set of index $0\VEV{a}$ are represented by ket or bra. In particular, we denote
\be
\begin{split}
\ket{\e^{[i]}}
&= \lrbrk{\frac{\e^{0\VEV{a}{[i]}}}{(2p+1)!}} = \bra{\e^{[i]}},\\
\ket{\e_{[i]}}
&= \lrbrk{-\frac{\e_{0\VEV{a}{[i]}}}{(2p)!}} = \bra{\e_{[i]}},
\end{split}
\ee
\be\label{ketdTketwi}
\begin{split}
\ket{dT} &= \lrbrk{dt\w dx^{\VEV{a}}} = \bra{dT},\\
\ket{w^{[i]}} &= \lrbrk{w^{0\VEV{a},[i]}} = \bra{w^{[i]}}.
\end{split}
\ee
The normalisation of $\ket{\e^{[i]}}$ and $\ket{\e_{[i]}}$ are introduced for the convenience of inner and outer products.
Quantities with two sets of index $0\VEV{a}$ are suppressed
and are considered as linear operators acting on bra or ket.
In particular,
\be
w = (w^{0\VEV{a},0\VEV{b}}).
\ee
Note that the indices of the form $[i]$ are not suppressed. For convenience, we will denote
\be
w^{[i][j]}\equiv w^{[i],[j]}.
\ee
The contraction of unsuppressed indices is defined as usual. For example, eq.\eq{paPtaspaP} can be written as
\be
\ket{\paPt} = -\ket{\e^{[i]}}(\paP)_{[i]}.
\ee
The contraction of suppressed indices are defined such that one of the indices is upper while the other is lower.
For example,
\be
\ipr{\e^{[i]}}{\e_{[j]}} =  -\ove{(2p)!(2p+1)!}\e^{0\VEV{a}[i]}\e_{0\VEV{a}[j]} = \d^{[i]}_{[j]},
\ee
\be
\ket{\e^{[i]}}\bra{\e_{[i]}}
=\lrbrk{ -\ove{(2p)!(2p+1)!}\e^{0\VEV{a}[i]}\e_{0\VEV{b}[i]}}
= \mathbbm{1},
\ee
\be
w\ket{\e_{[i]}} = \lrbrk{-\frac{w^{0\VEV{a},0\VEV{b}}\e_{0\VEV{b}[i]}}{(2p)!}},
\ee
whereas the quantity $w\ket{\e^{[i]}}$ is undefined because the suppressed indices are all upper indices, so they cannot be contracted.

The identity \eq{preiden-we} implies
\be\label{iden-we-2-matrix}
w\ket{\e_{[j]}}w^{[i][j]} = -\ket{\e^{[i]}} + \ket{w^{[j]}}\ipr{w^{[i]}}{\e_{[j]}},
\ee
\be\label{iden-we-3-matrix}
\ket{\e_{[i]}}\ipr{w^{[i]}}{\e_{[k]}}w^{{[j]}{[k]}}
=\ket{\e_{[k]}}\ipr{w^{[j]}}{\e_{[i]}}w^{{[i]}{[k]}}.
\ee
Let $m_{[i][j]}$ be the inverse of $w^{[i][j]},$ that is
\be
m_{[i][j]}w^{[j][k]} = \d_{[i]}^{[k]}.
\ee
With this, eq.\eq{iden-we-2-matrix}-\eq{iden-we-3-matrix}
are respectively equivalent to
\be\label{iden-we-2-matrix-alt}
w = -\ket{\e^{[i]}}m_{{[i]}{[j]}}\bra{\e^{[j]}} + \ket{w^{[j]}}m_{{[i]}{[k]}}\bra{\e^{[k]}}\ipr{w^{[i]}}{\e_{[j]}},
\ee
\be\label{iden-we-3-matrix-alt}
\ipr{w^{[j]}}{\e_{[k]}}m_{{[j]}{[i]}}
=\ipr{w^{[j]}}{\e_{[i]}}m_{{[j]}{[k]}}.
\ee

In order to write $Q$ and $R,$ it is convenient to work out the eigenforms of $\bar{*}$ of the form
\be\label{eigenform}
\begin{split}
\ket{dT} &+ W_{\pm}^T\ket{\e_{[i]}}dx^{[i]}\\
&\equiv
\bigg(dt\w dx^{\VEV{a}}
-\ove{(2p)!}\e_{0\VEV{b}[i]} W_{\pm}^{0\VEV{b},0\VEV{a}} dx^{[i]}
\bigg),
\end{split}
\ee
which correspond to eivenvalues $\pm 1.$
By writing eq.\eq{sbar-dT}-\eq{sbar-dX} in bra-ket notation as
\be
\bar{*}\ket{dT}
=- \ket{w^{[i]}}\ipr{\e_{[i]}}{dT}
+w\ket{\e_{[i]}}dx^{[i]},
\ee
\be
\bar{*}dx^{[i]}
=-w^{[i][j]}\ipr{\e_{[j]}}{dT} + \ipr{w^{[i]}}{\e_{[j]}}dx^{[j]},
\ee
and then using identities \eq{iden-we-2-matrix-alt}-\eq{iden-we-3-matrix-alt}, we obtain the following identities
\be\label{Wpm-iden}
\begin{split}
	W_{\pm}^T
	&= W_{\pm}\\
	&=\mp\ket{\e^{[i]}}m_{{[i]}{[j]}}\bra{\e^{[j]}}
	-\ket{w^{[i]}}m_{{[i]}{[j]}}\bra{\e^{[j]}},
\end{split}
\ee
\be
w^{{[i]}{[j]}} = 2\bra{\e^{[i]}}(W_- - W_+)^{-1}\ket{\e^{[j]}},
\ee
\be
\begin{split}
	\ket{w^{[i]}}
	&= -(W_+ + W_-)(W_- - W_+)^{-1}\ket{\e^{[i]}},
\end{split}
\ee
\be\label{wAB-2}
\begin{split}
	w = 2(W_+^{-1} - W_-^{-1})^{-1},
\end{split}
\ee
\be\label{Wpminv}
W_{\pm}^{-1}
=\pm w^{-1}-w^{-1}\ket{w^{[i]}}\bra{\e_{[i]}},
\ee
where $w^{-1}$ is the inverse operator of $w.$
These in turn imply identities which we will also often make use of:
\be
\bra{w^{[i]}}w^{-1}\ket{w^{[j]}}
-w^{[i][j]} = \bra{\e^{[i]}}w^{-1}\ket{\e^{[j]}},
\ee
\be\label{ew-iden}
\bra{\e^{[i]}} + (\bra{w^{[i]}} - \bra{\e^{[i]}})w^{-1}W_+
=0.
\ee

It can be shown that by using eq.\eq{Wpm-iden}, the set of differential forms presented in \eq{eigenform} is linearly independent. The set can then be used as a basis. The projectors to $\bar{*}$-self-dual and $\bar{*}$-anti-self-dual basis are
\be
\hlf(1+\bar{*}),\qquad
\hlf(1-\bar{*}),
\ee
respectively. They act on coordinate basis as
\be\label{projbardT}
\begin{split}
\hlf&(1\pm\bar{*})\ket{dT}\\
&=\pm W_{\mp}(W_- - W_+)^{-1}(\ket{dT} + W_{\pm}\ket{\e_{[i]}}dx^{[i]}),
\end{split}
\ee
\be\label{projbardX}
\begin{split}
\hlf&(1\pm\bar{*})dx^{[i]}\\
&=\mp\bra{\e^{[i]}}(W_- - W_+)^{-1}(\ket{dT} + W_{\pm}\ket{\e_{[j]}}dx^{[j]}).
\end{split}
\ee

It will also be useful to consider differential $(2p+1)$-forms which are self-dual or anti-self-dual with respect to standard metric $g.$ For this, we define
\be
v^{\m_1\cdots\m_{2p+1},\n_1\cdots\n_{2p+1}}
\equiv\sqrt{-\det(g)}g^{[\m_1|\n_1|}\cdots g^{\m_{2p+1}]\n_{2p+1}},
\ee
and $V_{\pm}$ such that
\be\label{eigenform-V}
\ket{dT} + V_{\pm}\ket{\e_{[i]}}dx^{[i]}
=\pm*(\ket{dT} + V_{\pm}\ket{\e_{[i]}}dx^{[i]}).
\ee

\section{Analysis of a general nonlinear Sen theory}\label{sec:analysis}

\subsection{Hamiltonian analysis of a general nonlinear Sen theory}\label{subsec:Hamgen}
In this subsection, we present Hamiltonian analysis of a general Sen theory, in which the self-duality condition in the physical sector can be nonlinear, while the unphysical sector has linear self-duality condition. Furthermore, the metrics $g_{\m\n}, \bar{g}_{\m\n}$ are general.
Apart from being technically more involved, the analysis will closely follow the idea of the case $\bar{g}_{\m\n} = \h_{\m\n}$ given in
\cite{Sen:2019qit}, in which the decoupling between physical and unphysical fields at the Hamiltonian level is shown.
However, for the case of general $\bar{g}_{\m\n},$ the Lagrangian should take a certain form in order for the decoupling to be realised. We derive this form in this subsection.

As explained in section \ref{sec:review},
the action of
a general Sen theory encodes the dynamics
of the fields $P$ and $Q$ along with external unphysical metric $\bar{g}_{\m\n}$ and can also include physical external fields $g, \Psi.$
Due to $\bar{*}$-self-duality of $Q,$ one may choose half of the components of $Q$ to appear in the Lagrangian without loss of generality. In particular, let us choose the components $Q_{0a_1\cdots a_{2p}},$ which will be denoted in bra-ket notation as $\ket{Q}$ or $\bra{Q}.$
Similarly, since $R$ is $\bar{*}$-anti-self-dual,
we choose to present only the components 
$R_{0a_1\cdots a_{2p}},$ which will be denoted in bra-ket notation as $\ket{R}$ or $\bra{R}.$
We then express
\be\label{Qinb}
Q = \frac{(-1)^p}{(2p)!}(\ipr{Q}{dT} +\bra{Q}W_+\ket{\e_{[i]}} dx^{[i]}),
\ee
\be\label{Rinb}
R = \frac{(-1)^p}{(2p)!}(\ipr{R}{dT} +\bra{R}W_-\ket{\e_{[i]}} dx^{[i]}).
\ee
Note that
\be\label{QiRi}
\begin{split}
Q_{[i]} &= (2p+1)\bra{Q}W_+\ket{\e_{[i]}},\\
R_{[i]} &= (2p+1)\bra{R}W_-\ket{\e_{[i]}}.
\end{split}
\ee
For convenience of the readers,
we would like to reiterate the following
notations which are previously defined in section \ref{sec:conventions}.
The multi-index notations are $\VEV{a}=(a_1\cdots a_{2p}),$
$[i] = (i_1\cdots i_{2p+1}),$ as defined at the beginning of subsection \ref{subsec:mindex}. The notation $\bra{dT}$ or $\ket{dT}$ represent $dt\w dx^{\VEV{a}},$
as in eq.\eq{ketdTketwi}. The quantities $W_\pm$ are defined through the eigenforms of $\bar{*}$ as in eq.\eq{eigenform}.
These quantities are related to $w^{[\m][\n]}$ through eq.\eq{Wpm-iden}-\eq{Wpminv}.
The quantities $w^{[\m][\n]}$ in turn are related to the unphysical metric $\bar{g}$ by eq.\eq{wMN}.

The Lagrangian given by eq.\eq{Sen-gen}
can be expressed using bra-ket notation as
\be\label{Sen-gen-braket}
\begin{split}
	\cL
	&=\ove{8}(2p+1)\bra{\paP}w\ket{\paP}
	+\ove{4}\ipr{\paP}{w^{[i]}}(\paP)_{[i]}\\
	&\qquad+\ove{8}\ove{2p+1}(\paP)_{[i]}(\paP)_{[j]} w^{{[i]}{[j]}}\\
	&\qquad
	-\hlf\ipr{Q}{\e^{[i]}}(\paP)_{[i]}
	+\hlf(2p+1) \bra{Q}W_+\ket{\paP}\\
	&\qquad+\cL_I(Q, g, \bar{g},\Psi),
\end{split}
\ee
with
\be
\begin{split}
\d_Q&\cL_I(Q,g,\bar{g},\Psi)\\
&=-\ove{2}(2p+1)\bra{\d Q}(W_+ - W_-)\ket{R(Q, g,\bar{g},\Psi)}.
\end{split}
\ee
The combination $H = Q - R$ describes degrees of freedom of physical chiral $(2p)$-form field. It should satisfy a nonlinear self-duality condition which only involves $H, g,$ and $\Psi.$

Let us now work out the Hamiltonian analysis.
The analysis in this subsection is a direct generalisation of the analysis given in \cite{Sen:2019qit} in which $\bar{g} = \h$ is imposed. Here, we work with the case where $\bar{g}$ is a general unphysical metric.

Conjugate momenta for $P_{\m_1\cdots\m_{2p}}$ are denoted
\be\label{conjmom-P}
\p^{\m_1\cdots\m_{2p}}
=\frac{\pa\cL}{\pa\pa_0 P_{\m_1\cdots\m_{2p}}}.
\ee
After a direct calculation, we obtain
\be\label{pA}
\begin{split}
	\ket{\p}
	&=\ove{4}(2p+1)w\ket{\paP}
	+\ove{4}(\paP)_{[i]}\ket{w^{[i]}}\\
	&\qquad+\hlf(2p+1) W_+\ket{Q},
\end{split}
\ee
and
\be
\p^{0a_1\cdots a_{2p-1}} = 0,
\ee
where $\ket{\p}$ represents $\p^{a_1\cdots a_{2p}}.$
Conjugate momenta for $Q_{0a_1\cdots a_{2p}}$ are

\be\label{conjmom-Q}
\begin{split}
	\p_Q^{a_1\cdots a_{2p}}
	=\frac{\pa\cL}{\pa\pa_0 Q_{0a_1\cdots a_{2p}}}
	=0.
\end{split}
\ee
After some calculations, and by defining
\be
\p_\pm^{a_1\cdots a_{2p}} = \p^{a_1\cdots a_{2p}} \pm \ove{4}\paPt^{a_1\cdots a_{2p}},
\ee
or
\be\label{pipm}
\ket{\p_{\pm}} = \ket{\p} \pm \ove{4}\ket{\paPt},
\ee
we obtain the total Hamiltonian
	\be\label{Ham-gen}
\begin{split}
	\cH
	&=\frac{1}{2p+1}\bra{\p_-}W_+^{-1}\ket{\p_-}
	-\frac{1}{2p+1}\bra{\p_+}W_-^{-1}\ket{\p_+}
	\\
	&\qquad-\ove{4}(2p+1)\bra{Q}(W_+ - W_-)W_-^{-1}W_+\ket{Q}\\
	&\qquad+\bra{\p_+} W_-^{-1}(W_+-W_-)\ket{Q}
	-\cL_I\\
	&\qquad+ 2p\pa_{a_1}\p^{a_1 a_2\cdots a_{2p}}P_{a_2\cdots a_{2p}0}\\
	&\qquad+\z_{a_1\cdots a_{2p-1}}\p^{0 a_1\cdots a_{2p-1}}
	+u_{a_1\cdots a_{2p}}\p_Q^{a_1\cdots a_{2p}},
\end{split}
\ee
where $\z_{a_1\cdots a_{2p-1}}$ and $u_{a_1\cdots a_{2p}}$ are Lagrange multipliers
which enforce the primary constraints $\p^{0 a_1\cdots a_{2p-1}}\approx 0$ and $\p_Q^{a_1\cdots a_{2p}}\approx 0.$ Next, requiring time derivative of the primary constraints to vanish gives rise to secondary constraints
\be\label{gauss}
\pa_{a_1}\p^{a_1\cdots a_{2p}}\approx 0,
\ee
\be\label{QRP}
-(2p+1)\bra{Q}W_+
+(2p+1)\bra{R}W_-
+2\bra{\p_+}
\approx 0.
\ee
Using $H = Q-R,$ we may express eq.\eq{QRP} alternatively as
\be\label{Hpi}
H_{[i]}\approx 2\ipr{\p_+}{\e_{[i]}}.
\ee
Next, by computing time derivative of secondary constraints, it turns out that there is no further constraint.

Classification of constraints suggests that
first-class constraints are $\p^{0 a_1\cdots a_{2p-1}}\approx 0$ and $\pa_{a_1}\p^{a_1\cdots a_{2p}}\approx 0,$
whereas second-class constraints are
$\p_Q^{a_1\cdots a_{2p}}\approx 0$
and $H_{[i]}\approx 2\ipr{\p_+}{\e_{[i]}}.$
After constraints are classified, the number of degrees of freedom can be determined. It turns out that the number of degrees of freedom for Sen action is equal to twice of the number of degrees of chiral $(2p)$-form fields. This is as expected since Sen theory contains two chiral $(2p)$-form fields, one of which is physical while the other is unphysical with the wrong sign of kinetic energy.

Next, one may then solve second-class constraints to reduce the number of phase space variables.
In particular, it is natural to eliminate $\p_Q^{a_1\cdots a_{2p}}$ and $Q_{0a_1\cdots a_{2p}}.$
This is by solving the second-class constraints for
$\p_Q^{a_1\cdots a_{2p}}$ and $Q_{a_1\cdots a_{2p}}$
and substituting into the Hamiltonian. Eliminating $\p_Q^{a_1\cdots a_{2p}}$ is straightforward.
As for eliminating $Q_{a_1\cdots a_{2p}},$
let us first use the definition $H = Q - R,$
which gives $\bra{H} = \bra{Q} - \bra{R}.$
So eq.\eq{QRP} gives
\be\label{QtoHpi}
\begin{split}
\bra{Q} \approx& -\bra{H}W_-(W_+ - W_-)^{-1}\\
&\quad+ \frac{2}{2p+1}\bra{\p_+}(W_+ - W_-)^{-1}
\end{split}
\ee
\be
\begin{split}
\bra{R} \approx& -\bra{H}W_+(W_+ - W_-)^{-1}\\
&\quad+ \frac{2}{2p+1}\bra{\p_+}(W_+ - W_-)^{-1}.
\end{split}
\ee
One may then eliminate $\bra{H}$ by using nonlinear self-duality condition to write it in terms of $H_{[i]}, g,$ and $\Psi$ then apply to eq.\eq{Hpi}.

The resulting Hamiltonian after the substitution of $\ket{Q}$ as a function of $\ket{\p_+}, g, \bar{g}, \Psi$ and after all the constraints are eliminated shows the separation between $\ket{\p_-}$ and $\ket{\p_+}$.
This means that the Hamiltonian is of the form
\be
\cH = \cH_- + \cH_+,
\ee
where
\be
\cH_- = \frac{1}{2p+1}\bra{\p_-}W_+^{-1}\ket{\p_-},
\ee
\be
\begin{split}
\cH_+
&= \cH_+(\ket{\p_+}, g, \bar{g}, \Psi)\\
&=-\frac{1}{2p+1}\bra{\p_+}W_-^{-1}\ket{\p_+}
\\
&\qquad-\ove{4}(2p+1)\bra{Q}(W_+ - W_-)W_-^{-1}W_+\ket{Q}\\
&\qquad+\bra{\p_+} W_-^{-1}(W_+-W_-)\ket{Q}
-\cL_I,
\end{split}
\ee
subject to $\pa_{a_1}\p^{a_1\cdots a_{2p}}\approx 0,$
and where $\bra{Q}$ is determined from eq.\eq{QtoHpi}
along with the nonlinear self-duality condition.
The Hamiltonian $\cH_-$ has the wrong sign. So $\ket{\p_-}$ is unphysical. Furthermore, $\ket{\p_-}$ only couples to $\bar{g}$ (through $W_+$).
As for the Hamiltonian $\cH_+,$
it is
a function of $\ket{\p_+}, g, \bar{g}, \Psi.$
In the case $\bar{g} = \h$ originally considered in \cite{Sen:2019qit}, $\cH_+$ only contains physical fields. So it looks as if any form of $\cL_I$ can be chosen so that the Hamiltonian separates into two disconnected sectors.
However, in the case where $\bar{g}$ is dynamical,
this metric is unphysical since it couples to 
unphysical mode $\ket{\p_-}.$
So $\cH_+$ should in fact be independent from $\bar{g},$ otherwise the unphysical and physical sector will be mixed.
In this case, the form of 
$\cL_I$ cannot be given arbitrarily.

Note that if one works in the framework in which $\bar{g}$ is fixed,
the form of $\cL_I$ is not restricted as there is no problem even in the case where the physical chiral field is coupled to $\bar{g}$. Therefore, the separation of the two sectors at the Hamiltonian level is achieved without the need to impose the conditions on
$\cL_I$. On the other hand, the restriction to the form of $\cL_I$
is only applicable in the framework in which $\bar{g}$ is dynamical.

In order to determine $\cL_I,$ let us note that by using eq.\eq{QRP}
$\cH_+$ can be expressed as
\be
\cH_+
=-\ove{4}(2p+1)\bra{Q}(W_+ - W_-)\ket{R}
-\hlf\ipr{\p_+}{H}
-\cL_I.
\ee
It is clear that $\cL_I$ should take the form
\be
\cL_I = \cL_I^{(1)}(Q, g, \bar{g},\Psi) + \cL_I^{(2)}(H, g, \Psi).
\ee
The term $\cL_I^{(1)}(Q, g, \bar{g},\Psi)$ should be such that $\cH_+$ is independent from $\bar{g}$ while the term $\cL_I^{(2)}(H, g, \Psi)$ only contains physical degrees of freedom.
In particular, one can set without the loss of generality,
\be\label{cLI_class}
\cL_I
=-\ove{4}(2p+1)\bra{Q}(W_+ - W_-)\ket{R}
+\cL_I^{(2)}(H, g, \Psi),
\ee
which can be expressed using differential form as
	\be\label{cLI_diff}
	\cL_I
	=\frac{(2p)!}{4}Q\w R
	+\cL_I^{(2)}(H, g, \Psi).
	\ee
In the case of linear self-duality and in the absence of source, we have $\cL_I^{(2)}(H, g, \Psi) = 0.$

Having determined the form of general Lagrangian,
we can now see that the Hamiltonian \eq{Ham-gen} takes the form, after all secondary constraints are eliminated
\be\label{H-result}
\begin{split}
	\cH
	&\approx\frac{1}{2p+1}\bra{\p_-}W_+^{-1}\ket{\p_-}
	-\hlf\ipr{\p_+}{H}\\
	&\qquad-\cL_I^{(2)}(H, g, \Psi)\\
	&\qquad+ 2p\pa_{a_1}\p^{a_1 a_2\cdots a_{2p}}P_{a_2\cdots a_{2p}0}\\
	&\qquad+\z_{a_1\cdots a_{2p-1}}\p^{0 a_1\cdots a_{2p-1}},
\end{split}
\ee
where $H$ is related to $\ket{\p_+}$ by $H_{[i]} \approx 2\ipr{\p_+}{\e_{[i]}}$ and $H_{0\VEV{a}} = H_{0\VEV{a}}(\p_+^{\VEV{b}}, g, \Psi),$ in which the latter is obtained from the nonlinear self-duality condition for $H$.

Dirac bracket relations are identified with Poisson bracket relations\footnote{A quick check is to consider first-order Lagrangian corresponding to the Hamiltonian \eq{H-result}. After eliminating all the constraints,
	the terms containing time derivative are $\p^{a_1\cdots a_{2p}}\dot{P}_{a_1\cdots a_{2p}}.$ By the procedure of Faddeev-Jackiw \cite{Faddeev:1988qp}, \cite{Jackiw:1993in}, one immediately sees that Dirac brackets are simply given by Poisson brackets.}
\be
\begin{split}
	&[\p_\pm^{a_1\cdots a_{2p}}(t,\vec{x}),\p^{b_1\cdots b_{2p}}_\pm(t,\vec{x}')]\\
	&\quad=\mp\ove{2(2p)!}\frac{\pa}{\pa x^i}\d^{(4p+1)}(\vec{x}-\vec{x}')\e^{{a_1\cdots a_{2p}}{b_1\cdots b_{2p}}i},\\
	&[\p_\pm^{a_1\cdots a_{2p}}(t,\vec{x}),\p^{b_1\cdots b_{2p}}_\mp(t,\vec{x}')] = 0.
\end{split}
\ee

It is useful to discuss the relationship between the Hamiltonian variables with Lagrangian variables. From the constraint \eq{Hpi}, we have
\be\label{pHJ}
(\p_+)^{a_1\cdots a_{2p}}
=\hlf\e^{0a_1\cdots a_{2p}i_1\cdots i_{2p+1}}H_{i_1\cdots i_{2p+1}}.
\ee
Or 
\be
(\p_+)^{a_1\cdots a_{2p}}
=\hlf\sqrt{-\det(g)}(*H)^{0a_1\cdots a_{2p}},
\ee
where indices of $*H$ are raised by $g.$
Next, we note that eq.\eq{Hs-defn} implies
\be
\begin{split}
\bra{H_{(s)}}
&=\bra{Q} + \hlf\bra{\paP}w W_+^{-1}\\
&\qquad+\ove{2p+1}\bra{\paPt}(W_- - W_+)^{-1}.
\end{split}
\ee
Then by using eq.\eq{pA} and identities \eq{Wpm-iden}-\eq{Wpminv},
we obtain
\be
\ket{\p_-} = \hlf(2p+1)W_+\ket{H_{(s)}},
\ee
or in index notation,
\be\label{pHs}
\p_-^{a_1\cdots a_{2p}}
=\hlf\sqrt{-\det(\bar{g})}(H_{(s)})^{0a_1\cdots a_{2p}},
\ee
where indices of $H_{(s)}$ are raised by $\bar{g}.$
The relationships \eq{pHJ} and \eq{pHs} are direct generalisation to the analysis of \cite{Andriolo:2020ykk}, as one would expect.

Given the Lagrangian of a Sen theory,
i.e. with the form of $\cL_I^{(2)}$ given,
the Hamiltonian can immediately be written down by using eq.\eq{H-result}. However, we still need to determine how $H_{0\VEV{a}}$ is related to $\p_+^{\VEV{b}}, g, \Psi,$
which should be obtained from self-duality condition of the given theory. This is the only task left to do for each given Sen theory. Explicit examples will be discussed in sections \ref{sec:quad}-\ref{sec:nl6d}.

\subsection{Duality and decoupling at Lagrangian level}\label{subsec:dualdecoup}
In Sen theory, the decoupling \cite{Sen:2015nph}, \cite{Sen:2019qit}, \cite{Andriolo:2020ykk} between physical and unphysical chiral $(2p)$-form fields can be shown at the level of Hamiltonian as well as equations of motion. The decoupling at the Lagrangian level, however, is left to be explicitly shown.
A hint that this could be possible comes from the result of dimensional reduction of Sen theory and its generalisations \cite{Sen:2019qit}, \cite{Andriolo:2020ykk}, \cite{Phonchantuek:2023iao}, \cite{Chakrabarti:2023czz} that unphysical and physical fields are decoupled in the Lagrangian of the reduced theories.

In this subsection, we will present the decoupling of 
physical and unphysical fields at the Lagrangian level.
We apply the standard approach 
\cite{Tseytlin:1990ar}, \cite{Sonnenschein:1988ug}
(see also \cite{Abreu:1998tc}), which makes use of field redefinition on phase space variables to decouple chiral $(2p)$-form fields.
In passing, this approach also makes a duality present in the action.

More explicitly, the approach of 
\cite{Tseytlin:1990ar}, \cite{Sonnenschein:1988ug} is introduced to separate a two-dimensional Klein-Gordon scalar to two chiral bosons. This approach can easily be extended to help recognising
a $(2p)$-form electrodynamics in $(4p+2)$-dimensional spacetime as a theory of two chiral $(2p)$-forms.

The $p = 0$ case can be explained as follows.
One starts by writing the first-order form of Klein-Gordon Lagrangian: $\cL = \p\pa_0{\f} - \p^2/2 - (\pa_1\f)^2/2,$ where $\f$ is the scalar field and $\p$ is the conjugate momentum of $\f.$
Then one makes a field redefinition $\f = \f_+ + \f_-, \p = \pa_1\f_+ - \pa_1\f_-$
giving $\cL = \pa_0{\f_+}\pa_1\f_+ - (\pa_1\f_+)^2 - \pa_0\f_-\pa_1\f_- - (\pa_1\f_-)^2.$
With this form, one easily sees the decoupling between
the chiral bosons $\f_+$ and $\f_-$.
The Dirac bracket relations for $\f_+$ and $\f_-$ can directly be obtained from the Poisson bracket relations for $\f$ and $\p.$ One has
$[\pa_x\f_\pm(t,x),\pa_{x'}\f_\pm(t,x')] = \pm\pa_x\d(x-x')/2$
and $[\pa_x\f_\pm(t,x),\pa_{x'}\f_\mp(t,x')] = 0,$
which take the expected form.

In fact, before making the field redefinition as explained above, one may replace $\p$ by $\pa_1\chi.$ The Lagrangian then becomes $\cL = \pa_1\chi\pa_0\f - (\pa_1\chi)^2/2 - (\pa_1\f)^2/2.$ The dualisation $\f\lrar\chi$ leaves this Lagrangian invariant up to total derivative terms. Note that the dualisation corresponds to $\f_+\to\f_+, \f_-\to-\f_-.$

Let us now turn to the Lagrangian \eq{Sen-gen} of general Sen theory. The Hamiltonian analysis as discussed in subsection \ref{subsec:Hamgen} suggests that 
each of the physical and unphysical sector contain a chiral
$(2p)$-form. However, the Hamiltonian of the unphysical chiral $(2p)$-form field has the wrong sign.
Nevertheless, the same idea of field redefinition still applies.
Let us start from the first-order Lagrangian corresponding to the total Hamiltonian \eq{H-result}:
\be
\begin{split}
\cL
&=\p^{a_1\cdots a_{2p}}\pa_0 P_{a_1\cdots a_{2p}}\\
&\qquad+2p\p^{0a_1\cdots a_{2p-1}}\pa_0 P_{0a_1\cdots a_{2p-1}}
-\cH.
\end{split}
\ee
By eliminating $\p^{0a_2\cdots a_{2p}}$ and $\z_{a_2\cdots a_{2p}}$ using their equations of motion, the first-order Lagrangian reduces to
\be\label{LT_pip_gen}
\begin{split}
	\cL
	&=\p^{a_1\cdots a_{2p}}\pa_0 P_{a_1\cdots a_{2p}}\\
	&\qquad
	-\frac{\bar{N}^i}{(2p)!}\e_{ia_1\cdots a_{2p}b_1\cdots b_{2p}}
	\p_-^{a_1\cdots a_{2p}}\p_-^{b_1\cdots b_{2p}}\\
	&\qquad+\frac{\bar{N}}{\sqrt{\det(\bar{\g})}}\bar{\g}_{a_1 b_1}\cdots\bar{\g}_{a_{2p}b_{2p}}\p_-^{a_1\cdots a_{2p}}\p_-^{b_1\cdots b_{2p}}\\
	&\qquad+\hlf\ipr{\p_+}{H} + \cL_I^{(2)}(H, g, \Psi),
\end{split}
\ee
such that
$H_{[i]} \approx 2\ipr{\p_+}{\e_{[i]}}$ and, due to the nonlinear self-duality condition, $H_{0\VEV{a}} = H_{0\VEV{a}}(\p_+^{\VEV{b}}, g, \Psi).$
Furthermore, 
$\p^{a_1 a_2\cdots a_{2p}}$ should satisfy
\be
\pa_{a_1}\p^{a_1 a_2\cdots a_{2p}} = 0,
\ee
which can be solved by setting
\be
\begin{split}
\p^{a_1a_2\cdots a_{2p}}
&= \ove{4}\widetilde{\pa\f}^{a_1a_2\cdots a_{2p}}\\
&\equiv-\frac{1}{4(2p)!}\pa_{[i}\f_{b_1\cdots b_{2p}]}\e^{a_1\cdots a_{2p}ib_1\cdots b_{2p}}.
\end{split}
\ee
This leads to
\be
\p_\pm^{\VEV{a}} = \ove{4}(\widetilde{\pa\f}^{\VEV{a}} \pm \paPt^{\VEV{a}}),
\ee
which can be substituted into the first-order Lagrangian \eq{LT_pip_gen} giving
\be\label{LT_pip_gen2}
\begin{split}
	\cL
	&=\ove{4}\widetilde{\pa\f}\pa_0 P_{a_1\cdots a_{2p}}\\
	&\quad
	-\frac{\bar{N}^i\e_{ia_1\cdots a_{2p}b_1\cdots b_{2p}}}{16(2p)!}
	(\widetilde{\pa\f} - \paPt)^{a_1\cdots a_{2p}}(\widetilde{\pa\f} - \paPt)^{b_1\cdots b_{2p}}\\
	&\quad+\frac{\bar{N}\bar{\g}_{a_1 b_1}\cdots\bar{\g}_{a_{2p}b_{2p}}}{16\sqrt{\det(\bar{\g})}}(\widetilde{\pa\f} - \paPt)^{a_1\cdots a_{2p}}(\widetilde{\pa\f} - \paPt)^{b_1\cdots b_{2p}}\\
	&\quad+\tilde{\cL}\lrbrk{\frac{\widetilde{\pa\f} + \paPt}{2}, g, \Psi},
\end{split}
\ee
where $\tilde{\cL}((\widetilde{\pa\f} + \paPt)/2, g, \Psi)$ is obtained from $\ipr{\p_+}{H}/2 + \cL_I^{(2)}(H, g, \Psi)$ after the appropriate substitutions.
The Lagrangian \eq{LT_pip_gen2}
is unchanged under the duality transformation $P\lrar\f$
which corresponds to $\p_+\to\p_+, \p_-\to-\p_-,$
which can in turn be identified with duality transformation $H\to H, H_{(s)}\to -H_{(s)}$
that is discovered in \cite{Lambert:2023qgs} by directly dualising $P$ on the original Sen action.

Let us now consider the field redefinitions
\be\label{Psep}
P_{a_1\cdots a_{2p}}
=\r_{a_1\cdots a_{2p}} + \s_{a_1\cdots a_{2p}},
\ee
\be\label{phisep}
\f_{a_1\cdots a_{2p}}
=\r_{a_1\cdots a_{2p}} - \s_{a_1\cdots a_{2p}},
\ee
with
\be
\widetilde{\pa\r}^{a_1\cdots a_{2p}}
\equiv-\frac{1}{(2p)!}\pa_{[i}\r_{b_1\cdots b_{2p}]}\e^{a_1\cdots a_{2p}ib_1\cdots b_{2p}},
\ee
\be
\widetilde{\pa\s}^{a_1\cdots a_{2p}}
\equiv-\frac{1}{(2p)!}\pa_{[i}\s_{b_1\cdots b_{2p}]}\e^{a_1\cdots a_{2p}ib_1\cdots b_{2p}}.
\ee
With this, the Lagrangian \eq{LT_pip_gen} becomes
\be\label{LT_pip_gen_3}
\begin{split}
	\cL
	&=-\ove{4}\widetilde{\pa\s}^{a_1\cdots a_{2p}}\dot\s_{a_1\cdots a_{2p}}\\
	&\qquad
	-\ove{4}\frac{\bar{N}^i}{(2p)!}\e_{ia_1\cdots a_{2p}b_1\cdots b_{2p}}
	\widetilde{\pa\s}^{a_1\cdots a_{2p}}\widetilde{\pa\s}^{b_1\cdots b_{2p}}\\
	&\qquad+\ove{4}\frac{\bar{N}}{\sqrt{\det(\bar{\g})}}\bar{\g}_{a_1 b_1}\cdots\bar{\g}_{a_{2p}b_{2p}}\widetilde{\pa\s}^{a_1\cdots a_{2p}}\widetilde{\pa\s}^{b_1\cdots b_{2p}}\\
	&\qquad+\ove{4}\widetilde{\pa\r}^{a_1\cdots a_{2p}}\dot\r_{a_1\cdots a_{2p}}+\tilde{\cL}(\widetilde{\pa\r}, g, \Psi),
\end{split}
\ee
which clearly shows the decoupling between the two sectors at the Lagrangian level. Furthermore, the 
unphysical chiral
$(2p)$-form field is described by Henneaux-Teitelboim Lagrangian \cite{Henneaux:1988gg} but with the wrong sign of kinetic term and only couples to the unphysical metric $\bar{g}_{\m\n}.$
We will show explicitly in section \ref{sec:quad} that for the case of quadratic Sen theory, the physical chiral $(2p)$-form field is also described by 
Henneaux-Teitelboim Lagrangian with the correct sign of kinetic term.

\subsection{Diffeomorphism invariances for nonlinear Sen theories}
As part of the analysis in subsection \ref{subsec:Hamgen}, we have obtained the form of the Lagrangian for nonlinear Sen theories from the requirement that the Hamiltonian should be separated into two terms describing the two separate sectors. Let us now discuss diffeomorphism invariances. It is shown by \cite{Hull:2023dgp} that there are two types of diffeomorphism invariances in quadratic Sen theory. The first type is called the $\z$-transformation. It acts as standard diffeomorphism only on the physical sector but does not act on the unphysical sector.
The second type is called the $\chi$-transformation, which acts as standard diffeomorphism only on the unphysical sector but does not act on the physical sector.
More precisely, at the Lagrangian level, each of the transformation rules does not completely separate the two sectors. There are also extra terms in the transformation rules described above but in such a way that these extra terms vanish on-shell. 
In this section, let us show that diffeomorphism invariances can also be obtained in nonlinear Sen theories.

Let us focus on the case where the only physical external fields are the physical metric $g$ and external $(2p+1)$-form source J. From eq.\eq{Sen-gen}-\eq{dQLI}, \eq{cLI_diff},
we have
\be\label{cL_nl}
\begin{split}
	\cL = \frac{(2p)!}{2}\bigg(&\ove{4}dP\w\bar{*} dP - Q\w dP + \hlf Q\w R + \hlf H\w J\\
	&\ + \frac{\sqrt{-\det(g)}U(H^J,g)}{4(2p+1)!}\bigg),
\end{split}
\ee
with
\be\label{dQcLI}
\begin{split}
	\d_Q\bigg(&\hlf Q\w R + \ove{4(2p+1)!}\sqrt{-\det(g)}U(H^J,g)\\
		&\ + \hlf H\w J\bigg)=\d Q\w R,
\end{split}
\ee
where we have set
\be\label{cLI_HJ}
\cL_I^{(2)}(H,g,J)
=\frac{\sqrt{-\det{(g)}}U(H^J,g)}{8(2p+1)} + \frac{(2p)!}{4} H\w J.
\ee
The nonlinear self-duality condition is of the form
\be\label{nlsds_HJ}
*H^J = \cV(H^J,g),
\ee
where $H^J=H+J$.
Note that the introduction of $(2p+1)$-form source terms into eq.\eq{cLI_HJ}-\eq{nlsds_HJ} is suggested by ($\z$-type) diffeomorphism invariance analysed in the quadratic Sen theory \cite{Sen:2015nph}, \cite{Sen:2019qit}, \cite{Andriolo:2020ykk}, as well as the extension to $p=1$ nonlinear Sen theory \cite{Vanichchapongjaroen:2020wza}.

By expanding the LHS of eq.\eq{dQcLI}
and rearranging, we obtain
\be
\cV^{\m_1\cdots\m_{2p+1}}\d H^J_{\m_1\cdots\m_{2p+1}}
=-\hlf\frac{\d U}{\d H^J_{\m_1\cdots\m_{2p+1}}}\d H^J_{\m_1\cdots\m_{2p+1}}.
\ee
Note that $U$ is a scalar function under the standard diffeomorphism. So by following \cite{Cederwall:1997gg}, one may compute $\d U$ by making use of the relationship between the numbers of $(H^J)$'s and $g$'s within $U$. This gives
\be
\begin{split}
\d U
&=\frac{2(2p+1)!}{\sqrt{-\det(g)}}H^J\w\d H^J\\
&\qquad+(2p+1)\cV^{\r}{}_{\m_2\cdots\m_{2p+1}}(H^J)^{\s\m_2\cdots\m_{2p+1}}\d g_{\r\s}.
\end{split}
\ee
By using an identity
\be\label{VVHH-iden}
\begin{split}
	\cV\w\d\cV &= -H^J\w\d H^J
-\ove{(2p)!}\sqrt{-\det(g)}\cV^\r{}_{\m_2\cdots\m_{2p+1}}\\
&\qquad\times (H^J)^{\s\m_2\cdots\m_{2p+1}}\d g_{\r\s},
\end{split}
\ee
we have
\be\label{deltaU}
\d U
=\frac{(2p+1)!}{\sqrt{-\det(g)}}(H^J\w\d H^J-\cV\w\d\cV).
\ee

Following the strategy of \cite{Hull:2023dgp}, let us note that the Lagrangian \eq{cL_nl} is invariant under the standard diffeomorphism $\d_\xi = \pounds_\xi.$ 
We only need to work out $\z$-transformation $\d_\z^{\textrm{zeta}}$. After obtaining the $\z$-transformation, the $\chi$-transformation can be obtained by $\d_\chi^{\textrm{chi}} = \d_\chi - \d_\chi^{\textrm{zeta}}.$

The rule for $\z$-transformation is

\be
\begin{split}
	&\d_\z^{\textrm{zeta}}Q = -\frac{1+\bar{*}}{2}d\d_\z^{\textrm{zeta}}P,\qquad
	\d_\z^{\textrm{zeta}}g_{\m\n} = \pounds_\z g_{\m\n},\\
	&\d_\z^{\textrm{zeta}}\bar{g}_{\m\n} = 0,\qquad
	\d_\z^{\textrm{zeta}}J = \pounds_\z J.
\end{split}
\ee
We are left to determine $\d_\z^{\textrm{zeta}}P$.
The key idea of the calculation is to
make use of the identities \eq{VVHH-iden}-\eq{deltaU} for $\d = \d_\z^{\textrm{zeta}}$ and $\d = \pounds_\z$
and the fact that $\d_\z^{\textrm{zeta}} g_{\m\n} = \pounds_\z g_{\m\n}.$
Using this idea, the result can be obtained quite simply.
Explicit analysis is shown as follows.

From the identities \eq{VVHH-iden}-\eq{deltaU}, we have
\be
\begin{split}
	\frac{2}{(2p)!}\d_\z^{\textrm{zeta}}\cL
	&=tot. + d\d_\z^{\textrm{zeta}} P\w\lrbrk{\hlf\bar{*}dP + Q}\\
	&\qquad+\d_\z^{\textrm{zeta}} Q\w(R - dP)
	+di_\z H\w H\\
	&\qquad- \ove{4}\cV\w\d_\z^{\textrm{zeta}}\cV - \ove{4}H^J\w\d_\z^{\textrm{zeta}}H^J\\
	&\qquad
	+ \hlf H^J\w H^J\\
	&\qquad+ \frac{\sqrt{-\det(g)}g^{\m\n}U}{8(2p+1)!}\pounds_\z g_{\m\n}.
\end{split}
\ee
We may eliminate $\cV\w\d_\z^{\textrm{zeta}}\cV$ by using the
identity \eq{VVHH-iden} for $\d = \d_\z^{\textrm{zeta}}-\pounds_\z,$
and the identity \eq{deltaU} for $\d = \pounds_\z.$
We obtain
\be
\begin{split}
	\cV\w\d_\z^{\textrm{zeta}}\cV
&=2H^J\w\pounds_\z H^J - H^J\w\d_\z^{\textrm{zeta}} H^J\\
&\qquad - \frac{\sqrt{-\det(g)}}{(2p+1)!}\pounds_\z U .
\end{split}
\ee
So
\be\label{dzetaL}
\begin{split}
	\frac{2}{(2p)!}\d_\z^{\textrm{zeta}}\cL
	&=tot.+d\d_\z^{\textrm{zeta}} P\w\lrbrk{\hlf\bar{*}dP + Q}\\
	&\qquad
	+\d_\z^{\textrm{zeta}} Q\w(R - dP)
	+ di_\z H\w H.
\end{split}
\ee
By using
\be
\d_\z^{\textrm{zeta}}Q = -\frac{1+\bar{*}}{2}d\d_\z^{\textrm{zeta}}P,
\ee
we have,
\be
\begin{split}
	\frac{2}{(2p)!}\d_\z^{\textrm{zeta}}\cL
	&=tot.+(d\d_\z P+di_\z H)\w H.
\end{split}
\ee
So one may set
\be
\d_\z^{\textrm{zeta}} P = -i_\z H.
\ee
This finishes the proof of diffeomorphism for nonlinear Sen theory.

Note that the rule for $\d_\z^{\textrm{zeta}} P$ we have derived in general case is unchanged from the quadratic case \cite{Sen:2015nph}, \cite{Sen:2019qit}, \cite{Andriolo:2020ykk}, \cite{Hull:2023dgp} and the nonlinear case in six dimensions \cite{Vanichchapongjaroen:2020wza}.

\subsection{Elimination of $Q$ at equation of motion level}\label{Qelim}
From subsection \ref{subsec:dualdecoup}, we have seen that when going into phase space and eliminating fields apart from $P_{a_1\cdots a_{2p}}, \p^{a_1\cdots a_{2p}},$ and external fields, the first-order Lagrangian has a duality which involves $P$ and $\p.$ Furthermore, the separation between physical and unphysical chiral fields is also present at the Lagrangian level.

While the duality $H\to H, H_{(s)}\to - H_{(s)}$ as well as decoupling between physical and unphysical fields are easily seen at the equation of motion level, a puzzle remains.
The separation at Hamiltonian and Lagrangian level provides a slightly puzzling feature that
the field $P$ appears as part of
each of physical and unphysical chiral fields.
This is different from the analysis at the equations of motion level as presented by \cite{Sen:2015nph}, \cite{Sen:2019qit}, \cite{Andriolo:2020ykk}, in which $P$ is only contained in unphysical chiral field but not in physical chiral field. More explicitly, $P$
is combined with $Q$ in a particular way to form an unphysical field $H_{(s)},$
which is decoupled from the physical field $H$ which does not contain $P.$

A resolution to this puzzle is to note that
the analysis at the equation of motion level as given in
\cite{Sen:2015nph}, \cite{Sen:2019qit}, \cite{Andriolo:2020ykk} does not eliminate $Q,$
whereas
the analysis at Hamiltonian and Lagrangian level involve eliminating $Q.$ So essentially, the analysis are not in parallel with one another.
One should try to eliminate $Q$ at the equation of motion level to see if $P$ is contained in both physical and unphysical chiral fields.
This can indeed be done. The analysis in the case of $g_{\m\n} = \bar{g}_{\m\n} = \h_{\m\n}$ and with linear self-duality is given in \cite{Andrianopoli:2022bzr}, \cite{Evnin:2022kqn}. In this subsection, we will extend the analysis to the case of general Sen theory.

Equations of motion for $P$ and $Q$ of the Lagrangian $\eq{Sen-gen}$ are
\be\label{dQP}
d\lrbrk{Q + \hlf\bar{*}dP} = 0,
\ee
\be\label{Rsoln}
R = \hlf(dP - \bar{*}dP).
\ee
Solutions to eq.\eq{dQP} in topologically trivial spacetime are of the form
\be\label{Qsoln}
Q + \hlf\bar{*}dP = d\xi,
\ee
where $\xi$ is a $(2p)$-form such that
\be
d\lrbrk{\xi + \hlf P} = \bar{*}d\lrbrk{\xi + \hlf P}.
\ee
So $\xi + P/2$ is the unphysical chiral $(2p)$-form which couples to the unphysical metric $\bar{g}$.
Next, from eq.\eq{Rsoln} and \eq{Qsoln},
we obtain
\be
Q - R
=d\lrbrk{\xi - \hlf P}.
\ee
Suppose that the self-duality condition takes the form
\be
*(Q-R) = \cV(Q-R, g, \Psi).
\ee
Then
\be\label{sdxi}
*d\lrbrk{\xi - \hlf P} = \cV\lrbrk{d\lrbrk{\xi - \hlf P}, g, \Psi}.
\ee
So $\xi - P/2$ is the physical chiral $(2p)$-form whose field strength satisfies self-duality condition, eq.\eq{sdxi}.
Also from eq.\eq{sdxi}, we see that 
$\xi - P/2$ couples to the standard metric $g$ and external physical fields $\Psi.$

From the analysis,
it can be seen that if $Q$ is eliminated at the equation of motion level, then both the physical $\xi - P/2$ and unphysical $\xi + P/2$ fields contain $P.$
In other words, like the separation at Lagrangian level, $P$ can be expressed as the sum $P = (P/2 - \xi) + (\xi + P/2)$
of physical and unphysical fields, cf. eq.\eq{Psep}.

\section{Separation of the two sectors of quadratic Sen theory}\label{sec:quad}
Although the analysis in the previous section provides some useful insights on the properties of a general form of Sen theory, in practice one would also work directly with explicit theories. As discussed above, what remains to be done in order to obtain the explicit final form of the Hamiltonian and Lagrangian for each specific theory is to solve the self-duality condition and determine the corresponding $\cL_I^{(2)}$.

In this section, we will consider the quadratic Lagrangian \eq{cL} and derive the
explicit form at the Hamiltonian and Lagrangian levels
showing the separations of the two sectors.
As a consistency check, we will also derive diffeomorphism rules from the Hamiltonian.

It is convenient to write $\cL_I^{(2)}$
for the quadratic Lagrangian \eq{cL}
as a function of $H^J = H + J.$
We have
\be\label{cL2_quad}
\cL_I^{(2)}
=\ove{4}\ipr{H^J}{\e^{[i]}}J_{[i]}
-\ove{4}\ipr{J}{\e^{[i]}}H^J_{[i]}.
\ee

\subsection{Decoupling at Hamiltonian level of quadratic Sen theory}
Self-duality condition $H^J = *H^J$ gives
\be\label{AtoI}
\bra{H^J}
=\ove{2p+1}H_{[i]}^J\bra{\e^{[i]}}V_+^{-1}.
\ee
We can relate $H_{[i]}^J$ with $\bra{\p_+}$
by using eq.\eq{Hpi}, which after introducing source is equivalent to
\be\label{Itopi}
H_{[i]}^J \approx 2\ipr{\p_+^J}{\e_{[i]}},
\ee
where
\be\label{pipJ}
\begin{split}
	\ket{\p^J_+}
	&\equiv\ket{\p_+} + \frac{2p+1}{2}(W_+\ket{J_+} + W_-\ket{J_-})\\
	&=\ket{\p_+} + \hlf J_{[i]}\ket{\e^{[i]}}.
\end{split}
\ee
Substituting eq.\eq{cL2_quad}-\eq{Itopi} into eq.\eq{H-result}, we obtain
\be\label{Ham_quad_red}
\begin{split}
	\cH
	&\approx\frac{1}{2p+1}\bra{\p_-}W_+^{-1}\ket{\p_-}
	-\frac{1}{2p+1}\bra{\p_+^J}V_+^{-1}\ket{\p_+^J}\\
	&\qquad+\ipr{\p_+^J}{J}
	-\frac{1}{4}J_{[i]}\ipr{\e^{[i]}}{J}\\
	&\qquad+ 2p\pa_{a_1}\p^{a_1 a_2\cdots a_{2p}}P_{a_2\cdots a_{2p}0}\\
	&\qquad
	+\z_{a_1\cdots a_{2p-1}}\p^{0 a_1\cdots a_{2p-1}}.
\end{split}
\ee

It is now useful to express the Hamiltonian in terms of index notation.
For this, let us express $W_{+}^{-1}$ and $V_+^{-1}$ in index notation using ADM variables.
Direct calculation gives
\be
\begin{split}
	&(W^{-1}_\pm)_{a_1\cdots a_{2p}b_1\cdots b_{2p}}\\
&\qquad=\mp\frac{\bar{N}}{\sqrt{\det(\bar{\g})}}(2p+1)
\bar{\g}_{[a_1 |b_1|}\cdots\bar{\g}_{a_{2p}]b_{2p}}\\
&\qquad\qquad
+\frac{2p+1}{(2p)!}\bar{N}^i\e_{a_1\cdots a_{2p} i b_1\cdots b_{2p}},
\end{split}
\ee
and
\be
\begin{split}
	&(V_\pm^{-1})_{a_1\cdots a_{2p}b_1\cdots b_{2p}}\\
&\qquad=\mp\frac{N}{\sqrt{\det(\g)}}(2p+1)
\g_{[a_1 |b_1|}\cdots\g_{a_{2p}]b_{2p}}\\
&\qquad\qquad+\frac{2p+1}{(2p)!}N^i\e_{a_1\cdots a_{2p} i b_1\cdots b_{2p}}.
\end{split}
\ee
The Hamiltonian then becomes
\be\label{HT_sep}
\begin{split}
	\cH
	&\approx\frac{\bar{N}^i}{(2p)!}\e_{ia_1\cdots a_{2p}b_1\cdots b_{2p}}
	\p_-^{a_1\cdots a_{2p}}\p_-^{b_1\cdots b_{2p}}\\
	&\qquad
	-\frac{\bar{N}}{\sqrt{\det(\bar{\g})}}\bar{\g}_{a_1 b_1}\cdots\bar{\g}_{a_{2p}b_{2p}}\p_-^{a_1\cdots a_{2p}}\p_-^{b_1\cdots b_{2p}}\\
	&\qquad-\frac{N^i}{(2p)!}\e_{ia_1\cdots a_{2p}b_1\cdots b_{2p}}
	(\p_+^J)^{a_1\cdots a_{2p}}(\p_+^J)^{b_1\cdots b_{2p}}\\
	&\qquad
	+\frac{N}{\sqrt{\det(\g)}}\g_{a_1 b_1}\cdots\g_{a_{2p}b_{2p}}(\p_+^J)^{a_1\cdots a_{2p}}(\p_+^J)^{b_1\cdots b_{2p}}\\
	&\qquad+(\p_+^J)^{a_1\cdots a_{2p}}J_{0a_1\cdots a_{2p}}\\
	&\qquad
	-\ove{4}\ove{(2p+1)!}J_{i_1\cdots i_{2p+1}}\e^{a_1\cdots a_{2p}i_1\cdots i_{2p+1}}J_{0a_1\cdots a_{2p}}\\
	&\qquad+2p\pa_{a_1}\p^{a_1 a_2\cdots a_{2p}}P_{a_2\cdots a_{2p}0}\\
	&\qquad + \z_{a_1\cdots a_{2p-1}}\p^{0 a_1\cdots a_{2p-1}}.
\end{split}
\ee

The explicit form of the Hamiltonian in the case of static background $g_{0i} = 0, \bar{g}_{\m\n} = \h_{\m\n}$ is presented in \cite{Andriolo:2020ykk}.
In our eq.\eq{HT_sep} we provide the generalisation to the case of general curved metrics $g_{\m\n}, \bar{g}_{\m\n}.$
Alternative to directly working out the Hamiltonian, one could also follow \cite{Andriolo:2020ykk} by expressing the Hamiltonian in terms of Lagrangian variables 
and then generalise the static spacetime case
to the case of general metric with the help of the identity
\be\label{HdnasHup}
\begin{split}
	H^J_{0a_1\cdots a_{2p}}
&=\frac{N\sqrt{\det(\g)}N^i}{(2p)!}\e_{ia_1\cdots a_{2p} b_1\cdots b_{2p}}(H^J)^{0b_1\cdots b_{2p}}\\
&\qquad
-N^2\g_{a_1 b_1}\cdots\g_{a_{2p}b_{2p}}(H^J)^{0b_1\cdots b_{2p}},
\end{split}
\ee
for $H^J = *H^J.$
In either of these methods, one would see that
the $\p_-$ and $\p_+^J$ part of the Hamiltonian separately coincide with Henneaux-Teitelboim Hamiltonian \cite{Henneaux:1988gg} for chiral $(2p)$-form. The explicit form of eq.\eq{HT_sep} suggests that the theory indeed describes two chiral $(2p)$-form with $\p_+$ corresponding to physical chiral field, while $\p_-$ corresponds to unphysical chiral field since its Hamiltonian has the wrong sign.

\subsection{Diffeomorphism of quadratic Sen theory from Hamiltonian analysis}
As a consistency check, let us attempt to derive diffeomorphism by using Hamiltonian.
In principle, one works with the total Hamiltonian defined on the full phase space consisting of all generalised coordinates and conjugate momenta. From the Hamiltonian,  one may obtain the diffeomorphism rules on all of the phase space variables.
After eliminating the conjugate momenta and Lagrange multipliers from the first-order Lagrangian corresponding to the total Hamiltonian, one then obtains the diffeomorphism rule on all the variables of the original Lagrangian.
We will first argue that this is too complicated and will show a workaround.

One may write the Hamiltonian of quadratic Sen theory in the full phase space by starting from eq.\eq{Ham-gen}.
Then let us define
\be\label{QJRJ}
Q^J = Q + J_+ ,\qquad
R^J = R - J_-,
\ee
\be\label{Jpm}
J_{\pm}
=\frac{J \pm \bar{*}J}{2}.
\ee
Then by using
$H^J = Q^J-R^J, H^J = *H^J, Q = \bar{*}Q, R = \bar{*}R,$
we obtain
\be
\begin{split}
\bra{H^J} &= \bra{Q^J} - \bra{R^J},\\
\bra{H^J}V_+ &= \bra{Q^J}W_+ - \bra{R^J}W_-,
\end{split}
\ee
which implies
\be
\bra{R^J} = \bra{Q^J}(V_+ - W_+)(V_+-W_-)^{-1}.
\ee
Substituting eq.\eq{cLI_class}, \eq{cL2_quad} into eq.\eq{Ham-gen} and using eq.\eq{QJRJ}-\eq{Jpm}, we obtain
the Hamiltonian in full phase space for quadratic Sen theory
\be\label{Ham-gen-lin}
\begin{split}
	\cH
	&=-\frac{1}{2p+1}\bra{\p_+^J}W_-^{-1}\ket{\p_+^J}
	+\frac{1}{2p+1}\bra{\p_-}W_+^{-1}\ket{\p_-}\\
	&\quad-\frac{(2p+1)}{4}
	\bra{Q^J}(W_- - W_+)W_-^{-1}V_+\\
	&\qquad\quad\times(V_+-W_-)^{-1}(W_- - W_+)\ket{Q^J}\\
	&\quad
	-\bra{\p_+^J} (W_+^{-1}-W_-^{-1}) W_+\ket{Q^J}\\
	&\quad-\frac{2p+1}{4}\bra{J_+}(W_+ - W_-)\\
	&\qquad\quad\times\lrbrk{\ket{J_-} + W_-^{-1}W_+\ket{J_+}}\\
	&\quad+ 2p\pa_{a_1}\p^{a_1 a_2\cdots a_{2p}}P_{a_2\cdots a_{2p}0}\\
	&\quad+\z_{a_1\cdots a_{2p-1}}\p^{0 a_1\cdots a_{2p-1}}
	+u_{a_1\cdots a_{2p}}\p_Q^{a_1\cdots a_{2p}}.
\end{split}
\ee
If we express the metric $g$ and $\bar{g}$ in terms of ADM variables, the complication would arise from the expression $(V_+ - W_-)^{-1}$ which mixes the two types of lapse functions and shift vectors. So the Hamiltonian depends on lapse functions and shift vectors in a complicated way, making it difficult to obtain the generators of diffeomorphism transformations.

Instead of working directly in the full phase space,
let us make an alternative attempt. In particular, we will work on the reduced phase space, on which the second-class constraints \eq{conjmom-Q}, \eq{QRP} vanish. The disadvantage of this approach is that the transformation rules of only some variables are obtained. However, it turns out that the Hamiltonian now is up to linear power in lapse functions and shift vectors, allowing the calculation to easily be carried out.
Making use of the second-class constraints on the Hamiltonian
\eq{Ham-gen-lin} gives rise to the Hamiltonian
\eq{Ham_quad_red}. The first-class constraints which generate diffeomorphisms are
\be
\cH_0 \equiv
\cH_0^{(g)} + \frac{\g_{a_1 b_1}\cdots\g_{a_{2p}b_{2p}}}{\sqrt{\det(\g)}}(\p_+^J)^{a_1\cdots a_{2p}}(\p_+^J)^{b_1\cdots b_{2p}},
\ee
\be
\cH_i \equiv 
\cH_i^{(g)}-\frac{\e_{ia_1\cdots a_{2p}b_1\cdots b_{2p}}}{(2p)!}
(\p_+^J)^{a_1\cdots a_{2p}}(\p_+^J)^{b_1\cdots b_{2p}},
\ee
\be
\bar{\cH}_0 \equiv
\cH_0^{(\bar{g})}-\frac{\bar{\g}_{a_1 b_1}\cdots\bar{\g}_{a_{2p}b_{2p}}}{\sqrt{\det(\bar{\g})}}\p_-^{a_1\cdots a_{2p}}\p_-^{b_1\cdots b_{2p}},
\ee
\be
\bar{\cH}_i \equiv
\cH_i^{(\bar{g})}
+\frac{\e_{ia_1\cdots a_{2p}b_1\cdots b_{2p}}}{(2p)!}
\p_-^{a_1\cdots a_{2p}}\p_-^{b_1\cdots b_{2p}},
\ee
where we have also included $g$ and $\bar{g}$ into the system of consideration. The Hamiltonian and momentum densities for these metrics are $\cH_0^{(g)}, \cH_i^{(g)}, \cH_0^{(\bar{g})}, \cH_i^{(\bar{g})}.$

The $\z$-transformation can be obtained from
\be
\d^{\textrm{zeta}}_\z\cdot
=[\cdot,\int d^{4p + 1}x (N\cH_0\z^0 + \cH_i(\z^i + \z^0 N^i))],
\ee
whereas the $\chi$-transformation can be obtained from
\be
\d^{\textrm{chi}}_\chi\cdot
=[\cdot,\int d^{4p + 1}x (\bar{N}\bar{\cH}_0\chi^0 + \bar{\cH}_i(\chi^i + \chi^0 \bar{N}^i))].
\ee
Diffeomorphism rules on some phase space functions will 
be the same after the reduced phase space is promoted to the full phase space. From a standard procedure (see for example exercises 3.16-3.17 in \cite{Henneaux:1994lbw}), these required phase space functions are coordinates on the second-class constraint surface such that each of the coordinates are chosen to Poisson commute with each of the second-class constraints.
We have
\be
\bra{\p_-},\qquad
\bra{P} - \frac{2}{2p+1}\bra{\p_Q}V_+^{-1}.
\ee
Next, since
$\p_Q\approx 0$ is a second-class constraint, we should have $\d\p_Q\approx 0.$ So we see that the diffeomorphism rules on $P$ is part of the symmetry of the Lagrangian \eq{cL}.
Note that although $\p_-$ could be identified with $H_{(s)}$ through eq.\eq{pHs}, the transformation rule 
$\d H_{(s)}$ is not necessarily identified with $\d \p_-.$

With this method, although one can only reliably obtain diffeomorphism rules on $g_{\m\n}, \bar{g}_{\m\n}, P,$
it is in fact possible to also determine diffeomorphism rules on $Q$ and $J.$ Since $J$ is a physical external field, we demand that $\d_\z^{\textrm{zeta}}J = \pounds_\z J, \d_\chi^{\textrm{chi}}J = 0.$ We are then left to determine diffeomorphism rules on $Q.$ Since it is the only unknown transformation, it can be obtained by demanding that the variation of the action vanishes.

For $\z$-transformation,
direct calculation gives, after combining with appropriate gauge transformation,
\be\label{zetaong}
\d^{\textrm{zeta}}_\z g_{\m\n}
=\pounds_\z g_{\m\n},
\ee
\be\label{zetaongbar}
\d^{\textrm{zeta}}_\z \bar{g}_{\m\n}
=0,
\ee
\be\label{zetaonP-0}
\d^{\textrm{zeta}}_\z P
=-i_\z H,
\ee
where as part of the process to obtain eq.\eq{zetaonP-0}, we have used the second-class constraint $\ipr{\p_+}{\e_{[i]}}\approx H_{[i]}/2$ to trade $\bra{\p_+}$ for $H_{[i]}.$
We can then determine $\d_\z^{\textrm{zeta}}Q$ by considering the $\z$-transformation of the Lagrangian \eq{cL}. This gives 
\be
\d_\z^{\textrm{zeta}}Q
=-\lrbrk{\frac{1+\bar{*}}{2}}d\d_\z^{\textrm{zeta}}P,
\ee
in agreement with
\cite{Andriolo:2020ykk}, \cite{Hull:2023dgp}.

For $\chi-$transformation, we obtain
\be\label{chiong}
\d^{\textrm{chi}}_\chi g_{\m\n}
=0,
\ee
\be\label{chiongbar}
\d^{\textrm{chi}}_\chi \bar{g}_{\m\n}
=\pounds_\chi \bar{g}_{\m\n},
\ee
\be\label{chionP-0}
\d^{\textrm{chi}}_\chi P
=i_\chi dP + i_\chi H,
\ee
where we have written $\bra{\p_-} = \bra{\p_+} - \bra{\paPt}/2$
then use $\ipr{\p_+}{\e_{[i]}}\approx H_{[i]}/2.$
The transformation rules \eq{chiong}-\eq{chionP-0} agree with part of the $\chi$-transformation rules derived in \cite{Hull:2023dgp}.

\subsection{Decoupling at Lagrangian level of quadratic Sen theory}
By applying the procedure of subsection \ref{subsec:dualdecoup}
to the Hamiltonain \eq{HT_sep},
we obtain
	\be\label{LT}
	\begin{split}
		\cL
		&=-\ove{4}\widetilde{\pa\s}^{a_1\cdots a_{2p}}\dot\s_{a_1\cdots a_{2p}}\\
		&\quad-\frac{\bar{N}^i\e_{ia_1\cdots a_{2p}b_1\cdots b_{2p}}}{4(2p)!}
		\widetilde{\pa\s}^{a_1\cdots a_{2p}}\widetilde{\pa\s}^{b_1\cdots b_{2p}}\\
		&\quad
		+\ove{4}\frac{\bar{N}}{\sqrt{\det(\bar{\g})}}\bar{\g}_{a_1 b_1}\cdots\bar{\g}_{a_{2p}b_{2p}}\widetilde{\pa\s}^{a_1\cdots a_{2p}}\widetilde{\pa\s}^{b_1\cdots b_{2p}}\\
		&\quad
		+
		\ove{4}(\widetilde{\pa\r} - \Jt)^{a_1\cdots a_{2p}}(\dot\r_{a_1\cdots a_{2p}} - J_{0a_1\cdots a_{2p}})\\
		&\quad+\frac{N^i\e_{ia_1\cdots a_{2p}b_1\cdots b_{2p}}}{4(2p)!}
		(\widetilde{\pa\r}-\Jt)^{a_1\cdots a_{2p}}(\widetilde{\pa\r}-\Jt)^{b_1\cdots b_{2p}}\\
		&\quad-\frac{N\g_{a_1 b_1}\cdots\g_{a_{2p}b_{2p}}}{4\sqrt{\det(\g)}}(\widetilde{\pa\r}-\Jt)^{a_1\cdots a_{2p}}(\widetilde{\pa\r}-\Jt)^{b_1\cdots b_{2p}}\\
		&\quad+\ove{4}\Jt^{a_1\cdots a_{2p}}(\dot{\r}_{a_1\cdots a_{2p}} - J_{0a_1\cdots a_{2p}})\\ &\quad-\ove{4}(\widetilde{\pa\r}-\Jt)^{a_1\cdots a_{2p}}J_{0a_1\cdots a_{2p}},
	\end{split}
	\ee
	which describes a theory with an unphysical chiral $(2p)$-form $\s$
	uncoupled with a physical chiral $(2p)$-form $\r.$
	The equation \eq{LT} shows explicitly that these chiral
	$(2p)$-form fields are described by Henneaux-Teitelboim Lagrangian \cite{Henneaux:1988gg} such that $\s$ has the wrong sign of kinetic term and that it is coupled only to the unphysical metric $\bar{g},$
	whereas 
	$\r$ has the correct sign of kinetic term and that it is coupled to the curved metric $g$ and source $J$.

\section{Explicit form of nonlinear Sen theory in six dimensions}\label{sec:nl6d}

\subsection{Generating nonlinear Sen theory in six dimensions}

In this subsection, we present a method to generate explicit nonlinear Sen theory in six dimensional spacetime.
In \cite{Vanichchapongjaroen:2020wza}, a similar problem in the case of $\bar{g}_{\m\n} = \h_{\m\n}$ is investigated
with the form of nonlinear self-duality condition suggested along with the corresponding explicit form of the action.
However, it was not clear whether the form obtained is general. Here, combined with the insight gained in subsection \ref{subsec:Hamgen}, we will justify that the form obtained in \cite{Vanichchapongjaroen:2020wza} is already general and can be naturally generalised to the case of general $\bar{g}_{\m\n}$.

For simplicity of discussion, let us consider the case where the only physical external field is the physical metric $g_{\m\n}.$ From eq.\eq{Sen-gen}-\eq{dQLI}, \eq{cLI_diff} The Lagrangian in the case $p=1$ reads
\be
\cL =\ove{4}dP\w\bar{*} dP - Q\w dP +
\hlf Q\w R + \cL_I^{(2)}(H,g),
\ee
with
\be\label{dQLI_2}
\d_Q\lrbrk{\hlf Q\w R + \cL_I^{(2)}(H,g)}
=\d Q\w R.
\ee
Let $H = Q - R$ satisfy nonlinear self-duality condition
\be\label{nlsds}
*H = \cV(H,g).
\ee
In order for the nonlinear self-duality condition to be consistent, $10$ equations out of the $20$ equations should be dependent on the other $10$ equations.
Our task is to explicitly determine the most general form of $\cV$ and determine the corresponding $\cL_I^{(2)}.$

Our strategy is to
determine $\cV$ and $\cL_I^{(2)}$
in the case where $g_{\m\n} = \h_{\m\n} = \bar{g}_{\m\n}.$ 
Then due to the covariance of these functions their dependence on $g$ can easily be restored.

In the case where $g_{\m\n} = \h_{\m\n} = \bar{g}_{\m\n},$
let us give more detailed explanation and justification to the relevant analysis given in \cite{Vanichchapongjaroen:2020wza}. Furthermore, some details might slightly differ.
In this case,
the only Lorentz covariant expression at the first power in $Q$ is
\be
Q_{\m\n\r},
\ee
which is totally antisymmetric and $*$-self-dual.
Next, by using the identity
\be
Q^{\m\n\r}Q_{\s\l\r} = \d^{[\m}_{[\s}Q^{\n]\r\t}Q_{\l]\r\t},
\ee
it can be seen that the only independent Lorentz covariant expression at the third power in $Q$ is
\be
(Q^3)_{\m\n\r} \equiv Q_{\m\n\r'}Q^{\m'\n'\r'}Q_{\m'\n'\r},
\ee
which is totally antisymmetric and $*$-anti-self-dual.
These quantities are related by
\be
(Q^2)^\n_\m Q_{\n\r\s} = (Q^3)_{\m\r\s},
\ee
where
\be
(Q^2)^\m_\n \equiv Q^{\m\r\s}Q_{\n\r\s}.
\ee
Due to self-duality of $Q_{\m\n\r}$ and Cayley-Hamilton theorem, the only independent Lorentz invariant expression is
\be
[Q^4] \equiv (Q^4)^\m_\m,
\ee
where
\be
(Q^4)^\m_\n \equiv (Q^2)^\m_\r(Q^2)^\r_\n.
\ee
Let us denote
\be\label{XQ4}
X \equiv \ove{6}[Q^4].
\ee
We have the following identities
\be
Q_{\n\r\s}(Q^3)^{\m\r\s} = (Q^4)^\m_\n,
\ee
\be
(Q^2)^{\n}_\m (Q^3)_{\n\r\s} = XQ_{\m\r\s},
\ee
\be
(Q^3)_{\n\r\s}(Q^3)^{\m\r\s}
=X(Q^2)^\m_\n.
\ee
Therefore, the Lorentz covariant expressions at $5\textrm{th}, 7\textrm{th}, 9\textrm{th},\cdots$ powers in $Q$ are expressible in terms of $Q_{\m\n\r}$ and $(Q^3)_{\m\n\r}.$
Therefore, without the loss of generality, $R_{\m\n\r}$ should take the form
\be\label{RfQ3}
R_{\m\n\r} = f(X)(Q^3)_{\m\n\r}.
\ee
Then,
\be
H_{\m\n\r} = Q_{\m\n\r} - f\cdot (Q^3)_{\m\n\r}.
\ee
By direct calculation, it can be seen that $(H^5)_{\m\n\r}$ is expressible in terms of $H_{\m\n\r}$ and $(H^3)_{\m\n\r}.$
Next, let us denote
\be
\begin{split}
Y_1&\equiv \ove{6}H_{\m\n\r}H^{\m\n\r},\\
Y_2&\equiv \ove{6}H_{\m\n\r}H^{\m\n\r'}H_{\m'\n'\r'}H^{\m'\n'\r}.
\end{split}
\ee
We have
\be\label{Y1Y2}
Y_1 = -2fX,\qquad
Y_2 = (1+6f^2 X + f^4 X^2)X.
\ee
It is clear that both $Y_1$ and $Y_2$ are functions of $X.$
So $Y_1$ and $Y_2$ must be related.
So the most general form of nonlinear self-duality condition is
\be\label{nl_sds_HH3}
(*H)_{\m\n\r} = F_1(Y_1) H_{\m\n\r} + F_2(Y_1) (H^3)_{\m\n\r}.
\ee
Since eq.\eq{RfQ3} is equivalent to eq.\eq{nl_sds_HH3},
there must be conditions between $f, F_1, F_2.$
These conditions can be obtained by substituting eq.\eq{RfQ3} into eq.\eq{nl_sds_HH3}
giving
\be\label{F1F2}
F_1 = \frac{1+6f^2 X + f^4 X^2}{1-f^4X^2},\qquad
F_2 = \frac{2f}{1-f^4X^2}.
\ee
By noting that $F_1$ and $F_2 Y_1$ are both functions of $f^2 X,$ we can eliminate $f^2 X$ to obtain
\be
F_1^2 + 3Y_1 F_1 F_2 + 2Y_1^2 F_2^2 = 1,
\ee
which is an alternative form of an expression given in \cite{Vanichchapongjaroen:2020wza}.

The nonlinear self-duality condition \eq{nl_sds_HH3} can then be generated by first specifying $f(X),$ then
invert $Y_1 = -2fX$ to write $X$ in terms of $Y_1$
and finally substitute into eq.\eq{F1F2}. For example, choosing $f = 0$ gives linear self-duality condition,
whereas choosing \cite{Vanichchapongjaroen:2020wza}
\be
f = \sum_{n=0}^\infty \frac{(-1)^{n+1}}{3n+2}{4n+1\choose n}\frac{X^n}{2^{6n+2}}
\ee
gives the self-duality condition corresponding to M5-brane.
In general, from eq.\eq{F1F2}, we can freely choose $f(X)$ as long as $f(X)\neq \pm 1/\sqrt{X}$.

The equation \eq{nl_sds_HH3} can easily be extended to the equation in curved $g_{\m\n}.$
Next, let us determine $\cL_I^{(2)}(H,g).$ It should be a Lagrangian density in the physical sector. So without the loss of generality, it should take the form
\be
\cL_I^{(2)}(H,g) = \ove{24}\sqrt{-\det(g)}U(Y_1).
\ee
Next, by expanding LHS of eq.\eq{dQLI_2},
we obtain
\be\label{Up}
\ove{24}\sqrt{-\det(g)}U'\d Y_1
=\hlf H\w\d H.
\ee
By using eq.\eq{nl_sds_HH3}, we have
\be
U = -3\int \d Y_1\lrbrk{F_1 - F_1'Y_1 + \frac{F_1F'_2}{F_2}Y_1}.
\ee
In \cite{Vanichchapongjaroen:2020wza}, the alternative expression for $U$ was given in which
$F_1$ and $F_2$ were taken to be functions of $Y_2$ as there are cases with $Y_1 = 0$. Here, however, we have expressed $F_1$ and $F_2$ as functions of $Y_1.$ So we should single out the cases where $Y_1 = 0.$
From eq.\eq{Y1Y2}, this is only possible when $f=0.$
So from eq.\eq{F1F2}, we have $F_1 = 1, F_2 = 0,$ which correspond to linear self-duality condition. In this case, eq.\eq{Up} implies $U' = 0.$ So we can set $U = 0.$
Our choice of taking $F_1$ and $F_2$ to be functions of $Y_1$ is to make it convenient to work with strictly nonlinear self-dual cases.

With the procedure explained in this subsection, the task for generating the Lagrangians for nonlinear Sen theories in six dimensional spacetime with general metrics $g_{\m\n}, \bar{g}_{\m\n}$ is complete.

\subsection{Separation of the two sectors of M5-brane action in Sen formulation}
A generalisation of Sen theory to M5-brane action is given in \cite{Vanichchapongjaroen:2020wza}. The generalisation is by nonlinearising the case $p=1$ in such a way that the nonlinear self-duality condition takes the expected form. The action is further extended to a complete M5-brane action in Green-Schwarz formalism by including appropriate couplings to background $11$-dimensional supergravity, where the background fields are put on-shell.
These fields appear in the M5-brane action through pullbacks.
In particular, we denote $J$ as the pullback of $3$-form gauge superfield from the target superspace,
whereas the field $C_6$ is the pullback of $6$-form dual to the $3-$form gauge superfield. Furthermore, the metric $g$ is induced due to the embedding of M5-brane worldvolume into the 11-dimensional target superspace.

It is straightforward to generalise
an M5-brane action given in \cite{Vanichchapongjaroen:2020wza} by
promoting the non-dynamical flat metric to an unphysical curved metric $\bar{g}$.
This results in the Lagrangian given by
	\be\label{M5-Sen}
	\begin{split}
		\cL
		&=\frac{3}{8}\bra{\paP}w\ket{\paP}
		+\ove{4}\ipr{\paP}{w^{[i]}}(\paP)_{[i]}\\
		&\qquad+\ove{24}(\paP)_{[i]}(\paP)_{[j]} w^{{[i]}{[j]}}\\
		&\qquad
		-\hlf\ipr{Q}{\e^{[i]}}(\paP)_{[i]}
		+\frac{3}{2} \bra{Q}W_+\ket{\paP}\\
		&\qquad-\frac{3}{4}\bra{Q}(W_+ - W_-)\ket{R}
		\\
		&\qquad-\ove{4}H^J_{[i]}\ipr{\e^{[i]}}{J}
		+\ove{4}J_{[i]}\ipr{\e^{[i]}}{H^J}\\
		&\qquad+\ove{24}\sqrt{-\det(g)}U_{\textrm{M5}} + \hlf C_6,
	\end{split}
	\ee
where $\ket{R}$ is defined through
\be
\d_Q\cL_I
= -\frac{3}{2}\bra{\d Q}((W_+ - W_-)\ket{R},
\ee
\be\label{HJ}
H^J = Q - R + J,
\ee
and
\be\label{nl_sds}
\begin{split}
(*H^J)_{\m\n\r}
&= \lrbrk{-\frac{U_{\textrm{M5}}}{12} + \frac{24}{U_{\textrm{M5}}}}H^J_{\m\n\r}\\
&\qquad + \frac{6}{U_{\textrm{M5}}}H^J_{[\m|\n'\r'}(H^J)^{\m'\n'\r'}H^J_{\m'|\n\r]},
\end{split}
\ee
with
\be
\begin{split}
\cL_I
&=-\frac{3}{4}\bra{Q}(W_+ - W_-)\ket{R}
-\ove{4}H^J_{[i]}\ipr{\e^{[i]}}{J}\\
&\qquad+\ove{4}J_{[i]}\ipr{\e^{[i]}}{H^J}
+\ove{24}\sqrt{-\det(g)}U_{\textrm{M5}},
\end{split}
\ee
\be
U_{\textrm{M5}} = -24\sqrt{1+ \frac{H^J_{\m\n\r}(H^J)^{\m\n\r}}{24}},
\ee
where indices of $H^J$ are raised by $g^{\m\n}.$ The symmetry corresponding to supergauge transformation and the kappa symmetry of the Lagrangian \eq{M5-Sen} still holds
such that the symmetry variation of each symmetry can be straightforwardly generalised by including $\d \bar{g}_{\m\n} = 0.$

The expression for conjugate momenta for $P$ and $Q$ are still given by eq.\eq{conjmom-P}-\eq{conjmom-Q} with $p = 1$. So the primary constraints are still the same as the quadratic action. The total Hamiltonian becomes
	\be\label{HamM5}
	\begin{split}
		\cH
		&=-\frac{1}{3}\bra{\p_+}W_-^{-1}\ket{\p_+}
		+\frac{1}{3}\bra{\p_-}W_+^{-1}\ket{\p_-}\\
		&\qquad-\frac{3}{4}\bra{Q}(W_+ - W_-)W_-^{-1}W_+\ket{Q}\\
		&\qquad+\frac{3}{4}\bra{Q}(W_+ - W_-)\ket{R}\\
		&\qquad+\ove{4}H^J_{[i]}\ipr{\e^{[i]}}{J}
		-\ove{4}J_{[i]}\ipr{\e^{[i]}}{H^J}\\
		&\qquad+\bra{\p_+} W_-^{-1}(W_+-W_-)\ket{Q}\\
		&\qquad-\ove{24}\sqrt{-\det(g)}U-\hlf C_6\\
		&\qquad+ 2\pa_{a_1}\p^{a_1 a_2}P_{a_20}
		+\z_{a}\p^{0 a}
		+u_{a_1a_{2}}\p_Q^{a_1a_2},
	\end{split}
	\ee
where $\ket{\p_{\pm}}$ is as defined in eq.\eq{pipm}.
Secondary constraint from $\dot{\p}^{0a} \approx 0$ are
\be
\pa_{a_1}\p^{a_1a_{2}}\approx 0.
\ee
Secondary constraint from $\dot{\p}^{ab}_Q \approx 0$ are
\be\label{condpQ-nl}
-3\bra{Q}W_+
+3\bra{R}W_-
+2\bra{\p_+}
\approx 0,
\ee
or in index-notation,
\be
-Q^{0ab} - R^{0ab} + 2\p_+^{ab} \approx 0,
\ee
where indices of $Q$ and $R$ are raised by $\bar{g}.$
Next, computing time derivative of secondary constraints give no further constraints.

Let us solve eq.\eq{condpQ-nl} for $\ket{Q}$ and substitute into the Hamiltonian. Instead of direct calculation, we may do this in steps. First, we solve eq.\eq{condpQ-nl} and \eq{HJ} to obtain $\ket{Q},$ $\ket{R},$ and $H_{[i]}^J$ in terms of $\ket{\p_+},$ $\ket{H^J},$ and $J.$
We have
\be
\bra{Q} \approx -\bra{H}W_-(W_+ - W_-)^{-1} + \frac{2}{3}\bra{\p_+}(W_+ - W_-)^{-1},
\ee
\be
\bra{R} \approx -\bra{H}W_+(W_+ - W_-)^{-1} + \frac{2}{3}\bra{\p_+}(W_+ - W_-)^{-1},
\ee
\be\label{HJpi}
H^J_{[i]} \approx 2\ipr{\p_+^J}{\e_{[i]}}.
\ee
After substituting into the Hamiltonian \eq{HamM5}, we obtain
	\be
\begin{split}
	\cH
	&=
	\frac{1}{3}\bra{\p_-}W_+^{-1}\ket{\p_-}\\
	&\qquad
	+\ipr{\p_+^J}{J}
	-\hlf\ipr{H^J}{\p_+^J}
	-\ove{4}J_{[i]}\ipr{J}{\e^{[i]}}\\
	&\qquad-\ove{24}\sqrt{-\det(g)}U-\hlf C_6\\
	&\qquad+ 2\pa_{a_1}\p^{a_1 a_2}P_{a_20}
	+\z_{a}\p^{0 a},
\end{split}
\ee
where $\ket{\p_+^J}$ is as defined in eq.\eq{pipJ}.
At this stage, $Q$ appears in the Hamiltonian only through $H^J.$
The idea is then to solve the self-duality condition for $H^J$ and substitute into the Hamiltonian.
For this,
we note that eq.\eq{nl_sds} is equivalent to
\be\label{nl_sds_2}
(H^J)^0{}_a{}^b
=\frac{(1+z_1)(\Ht^J)_a{}^b - \ove{g^{00}}((\Ht^J)^3)_a{}^b}{\sqrt{1+ z_1 + \hlf z_1^2 - z_2}},
\ee
where $\Ht^J$ is a $5\times 5$ matrix with components
\be\label{Htdef}
(\Ht^J)_a{}^b = \ove{6}\ove{\sqrt{-\det(g)}}\e^{0mbijk} H_{ijk}^Jg_{ma},
\ee
and
\be
z_1\equiv \hlf\ove{g^{00}} [(\Ht^J)^2],\qquad
z_2\equiv \ove{4}\ove{(g^{00})^2} [(\Ht^J)^4],
\ee
with $[\cdot]$ being the trace of a matrix.
By using eq.\eq{HJpi}, we may realise $z_1$ and $z_2$ as
\be
z_1 = -\frac{2}{\g}[(\p_+^J)^2],\qquad
z_2 = \frac{4}{\g^2}[(\p_+^J)^4],
\ee
with $\p_+^J$ being a $5\times 5$ matrix with components
\be
(\p_+^J)_a{}^b = \g_{ai}(\p_+^J)^{ib}.
\ee
By using eq.\eq{nl_sds_2}, we obtain
\be
U = -\frac{12(2+z_1)}{\sqrt{1+ z_1 + \hlf z_1^2 - z_2}},
\ee
and
\be\label{QmR-nl}
\begin{split}
H^J_{0ka}
&=-H^J_{iak} N^i\\
&\quad+N^2\g_{bk}\frac{(1+z_1)(\Ht^J)_a{}^b - \ove{g^{00}}((\Ht^J)^3)_a{}^b}{\sqrt{1+ z_1 + \hlf z_1^2 - z_2}}.
\end{split}
\ee
The Hamiltonian then reads
	\be
	\begin{split}
		\cH
		&=
		\frac{1}{3}\bra{\p_-}W_+^{-1}\ket{\p_-}\\
		&\quad
		-\hlf\e_{abimn}(\p_+^J)^{ab}(\p_+^J)^{mn}N^i\\
		&\quad+N\sqrt{\det(\g)}\sqrt{1+z_1+\hlf z_1^2 - z_2}\\
		&\quad+(\p_+^J)^{a_1 a_{2}}J_{0a_1 a_{2}}
		-\ove{24}J_{i_1 i_2 i_3}\e^{a_1a_2i_1 i_2 i_{3}}J_{0a_1a_{2}}\\
		&\quad
		-\hlf C_6
		+ 2p\pa_{a_1}\p^{a_1 a_2}P_{a_20}
		+\z_{a}\p^{0 a}.
	\end{split}
	\ee

Having obtained the Hamiltonian which shows the decoupling between physical and unphysical fields,
we may then proceed to obtain the Lagrangian
and use the field redefinition to show the decoupling.
By following the procedure of subsection \ref{subsec:dualdecoup}, this can indeed be achieved.
The resulting Lagrangian is
\be\label{cL-M5}
\begin{split}
	\cL
	&=-\ove{4}\widetilde{\pa\s}^{a_1 a_{2}}\dot\s_{a_1 a_{2}}
	-\ove{4}\frac{\bar{N}^i}{(2p)!}\e_{ia_1 a_{2}b_1 b_{2}}
	\widetilde{\pa\s}^{a_1 a_{2}}\widetilde{\pa\s}^{b_1 b_{2}}\\
	&\qquad+\ove{4}\frac{\bar{N}}{\sqrt{\det(\bar{\g})}}\bar{\g}_{a_1 b_1}\bar{\g}_{a_{2}b_{2}}\widetilde{\pa\s}^{a_1 a_{2}}\widetilde{\pa\s}^{b_1 b_{2}}\\
	&\qquad
	-\sqrt{-\det(g)}\ove{4}\frac{(*F)^{0a_1 a_{2}}F^0{}_{a_1a_2}}{g^{00}}\\
	&\qquad
	-\sqrt{-\det(g)}\sqrt{\det\lrbrk{\d_a^b + \ove{\sqrt{-g^{00}}}(*F)^0{}_a{}^b}}\\
	&\qquad+\hlf F\w J+\hlf C_6,
\end{split}
\ee
where we have defined
\be
F_{\m\n\r} = - 3\pa_{[\m}\r_{\n\r]} + J_{\m\n\r},
\ee
and correspondingly
\be
(*F)^{0ab} = \ove{\sqrt{-\det(g)}}(\widetilde{\pa\r}-\Jt)^{ab},
\ee
and that indices of $F$ and $*F$ are raised and lowered by $g.$

The action \eq{cL-M5} is the M5-brane action in gauge-fixed PST formulation \cite{Pasti:1997gx}, \cite{Bandos:1997ui}, \cite{Schwarz:1997mc} uncoupled from an unphysical chiral $2-$form field $\s$ and the unphysical metric $\bar{g}.$ This result shows that M5-brane action in Sen formulation can be put in the form of
the M5-brane action in gauge-fixed PST formulation.

\section{Conclusion and Discussion}\label{sec:conclusion}
In this work, we have studied aspects of nonlinear chiral $(2p)$-form theory in the Sen formulation \cite{Sen:2015nph}, \cite{Sen:2019qit}, \cite{Andriolo:2020ykk}, \cite{Hull:2023dgp}. The scope of the theories we investigate is as follows.
At the Lagrangian level, the theory encodes the dynamics of a $(2p)$-form $P$ and a $(2p+1)$-form $Q.$ Any external physical field can be introduced. In particular, the standard metric $g$ is introduced as part of external physical fields. As for external unphysical field, only the unphysical metric $\bar{g}$ is introduced.
The field strength of the unphysical chiral $(2p)$-form field is always linear self-dual with respect to $\bar{g}$,
whereas the field strength of the physical 
chiral $(2p)$-form field can be nonlinear self-dual
with respect to $g$.
We consider the framework
in which $g$ and $\bar{g}$ can both be promoted to become dynamical.
However, we focus on the systems of the fields $P$ and $Q$
while the other fields for example $g$ and $\bar{g}$ are external.

In particular, we work out the separation of the physical and unphysical sectors at the Hamiltonian and Lagrangian level. We started by considering the general Sen theories within the scope explained above. This gives properties in which these theories have in common.

We have shown that in order for the physical and unphysical sectors of a general Sen theory to be separated at the Hamiltonian level, the Lagrangian should take the form of eq.\eq{Sen-gen} with $\cL_I$ given by eq.\eq{cLI_diff}. Furthermore, the Hamiltonian for the chiral $(2p)$-form in the unphysical sector is described by the Henneaux-Teitelboim Hamiltonian \cite{Henneaux:1988gg} but with the incorrect overall sign.

The decoupling between the physical and unphysical sectors at the Lagrangian can also be shown. The idea is similar to the separation at the Lagrangian level of a 2D Klein-Gordon into two chiral scalar fields. That is, one considers the first-order Lagrangian of a general Sen theory. By eliminating all of the conjugate momenta and constraints
followed by making field redefinition \eq{Psep}-\eq{phisep},
we obtain the Lagrangian \eq{LT_pip_gen_3}.
It turns out that the unphysical chiral $(2p)$-form field is described by Henneaux-Teitelboim Lagrangian \cite{Henneaux:1988gg} but with the wrong sign of the kinetic term.

An example usage of our results is as follows. If one is given a nonlinear Sen theory
whose Lagrangian is described by eq.\eq{Sen-gen}, \eq{cLI_diff}
with $\cL_I^{(2)}$ specified, the separation of the two sectors at the Hamiltonian level and at the Lagrangian level are given by eq.\eq{H-result} and eq.\eq{LT_pip_gen_3}
respectively. For each theory, the only remaining task to do to obtain
the explicit form of these Hamiltonian and Lagrangian
is to solve the nonlinear self-duality condition for $H_{0\VEV{a}}$ in terms of $H_{[i]}.$

Two remarks are in order.
The first remark is that if one considers equations of motion for $P$ and $Q$
and eliminates $Q$ from the equations of motion,
then one can see that the physical chiral field is coupled to $g$
while the unphysical chiral field is coupled to $\bar{g}$, and hence the two sectors are separated. So there is no problem with $\bar{g}$ being dynamical.
Nevertheless, this is not yet a complete verification.
For this, one needs to study the full system whose Lagrangian encodes the dynamics of $P, Q, g, \bar{g}$. Then one should work out the separation at Lagrangian, or Hamiltonian, or equation of motion level. We leave this full verification as a future work.

The second remark is that as suggested by \cite{Hull:2023dgp},
there is another framework in which $\bar{g}$ is fixed.
In this case, the separation of the two sectors is achieved
without the need to restrict the form of $\cL_I$.

We then proceeded to derive the diffeomorphism invariance of general nonlinear Sen theories with general metrics $g, \bar{g}$ at the Lagrangian level. Even with this general consideration, the derivation can be done quite simply largely due to the use of the identities \eq{VVHH-iden}-\eq{deltaU}.
It turns out, as expected, that the $\z$-transformation rule on $P$ and $Q$ is unmodified from the quadratic theory \cite{Sen:2015nph}, \cite{Sen:2019qit}, \cite{Andriolo:2020ykk}, \cite{Hull:2023dgp} and the nonlinear theories in six dimensions \cite{Vanichchapongjaroen:2020wza}.

Having worked out common features of the general Sen theories, we then considered explicit example theories to obtain properties specific to each theory. We analysed the quadratic Sen theory whose Lagrangian is \eq{cL} and found that the Hamiltonian and Lagrangian of the physical chiral $(2p)$-form are described by the Henneaux-Teitelboim Hamiltonian and Lagrangian. We also considered the M5-brane action in the Sen formulation \cite{Vanichchapongjaroen:2020wza} and showed that the physical chiral $2$-form is described by the M5-brane action in the gauge-fixed PST formulation.
These results suggest that
at the Lagrangian and Hamiltonian levels, the Sen formulation is related to the gauge-fixed PST formulation.
This relationship at the Lagrangian level has not been discussed before in the literature. It is quite non-trivial to expect this since the roles of on-shell and off-shell conditions of the physical self-dual form are interchanged between
these two formulations.

As a consistency check, we have also worked out the diffeomorphism rules from the Hamiltonian of the quadratic Sen theory. The derivation can easily be carried out in the
reduced phase space since the Hamiltonian is up to linear powers in lapse functions and shift vectors. As a future work, it would be interesting to see whether it is possible to obtain the diffeomorphism rules in the full phase space on which the Hamiltonian is a complicated function of 
lapse functions and shift vectors.

We have also described explicitly how to generate nonlinear Sen theory in the case of $p = 1$. The upshot is that given any function $f(X)\neq \pm 1/\sqrt{X},$ where $X$ is defined in eq.\eq{XQ4}, there is a corresponding nonlinear Sen theory.
It would be interesting to generalise to the cases of other values of $p$.

Back to the relationship between the Sen formulation and the gauge-fixed PST formulation. There are several interesting directions for future investigations.
Let us start from noting that
the relationship between the two formulations might not be totally unexpected. Consider dimensional reduction of Sen theories \cite{Sen:2015nph}, \cite{Sen:2019qit}, \cite{Andriolo:2020ykk}, \cite{Phonchantuek:2023iao}, \cite{Chakrabarti:2023czz}. It is found for example that \cite{Andriolo:2020ykk}, \cite{Phonchantuek:2023iao} double dimensional reduction of M5-brane action in Sen formulation naturally gives rise to D4-brane or dual D4-brane action in the standard DBI form.
Double dimensional reduction of M5-brane action in gauge-fixed PST formulation also naturally gives rise to the dual D4-brane action in the standard DBI form.
Since the dimensionally reduced theories agree, one would seek for the relationship between the full theories. The results we obtained in this paper thus confirms the relationship.

From the previous paragraph, we have also seen that the M5-brane theory in Sen formulation also naturally gives rise to the D4-brane action. On the other hand, 
the M5-brane action within the gauge-fixed PST formulation giving rise naturally to standard D4-brane upon double dimensional reduction is the M5-brane action in the dual formulation \cite{Ko:2016cpw}
which is obtainable by gauge fixing the PST formulation with $5$ auxiliary scalar fields \cite{Ko:2017tgo} (see also \cite{Maznytsia:1998xw}, \cite{Maznytsia:1998yr}).
Therefore, it would be interesting to investigate as a future work to see whether the Sen formulation of chiral forms is also related at the Lagrangian level to the dual of the standard gauge-fixed PST formulation.

In fact, there are also alternative forms of gauge-fixed PST Lagrangians obtainable from different ways that the index for the full $(4p+2)$-dimensional spacetime is separated \cite{Chen:2010jgb}, \cite{Huang:2011np}. The relationships between these different forms at the Lagrangian level are not known. So 
it would be interesting to see if the Lagrangian in the Sen formulation also gives rise to these alternative Lagrangians.
If this is the case, it would mean that the Sen formulation connects the alternative forms of the gauge-fxied PST formulation at the Lagrangian level. Future investigations are required to see whether this is the case.

Looking further ahead, it would also be interesting to see whether the Sen formulation is directly related at the Lagrangian level to the standard PST formulation \cite{Pasti:1995ii}, \cite{Pasti:1996va}, \cite{Pasti:1996vs}, \cite{Pasti:1997gx}, \cite{Bandos:1997ui} and its alternatives \cite{Ko:2017tgo}, \cite{Maznytsia:1998xw}, \cite{Maznytsia:1998yr}, \cite{Pasti:2009xc}, \cite{Ko:2013dka}. It can then be hoped that if this is successful, one would obtain more insights into solving some problems for example the difficulties in constructing the $2+4$ formulation of M5-brane action \cite{Ko:2015zsy}.

It is also interesting to see whether the Sen formulation is related at the Lagrangian level to the clone field formulation \cite{Evnin:2022kqn}, \cite{Mkrtchyan:2019opf}, \cite{Avetisyan:2022zza}, \cite{Bansal:2023pnr}.

\section*{Acknowledgements}
We would like to thank Sheng-Lan Ko
for discussions. S.J. is supported by the Research Professional 
Development Project Under the Science Achievement 
Scholarship of Thailand (SAST). A.P. is supported by a Development and Promotion of Science and Technology Talents Project (DPST) scholarship from the Royal Thai Government.
This research by S.J., A.P., and P.V. has received funding support from the NSRF via the Program Management Unit for Human Resources \& Institutional Development, Research and Innovation [grant number B39G670016].

This version of the article has been (matched with the version)
accepted for publication, after peer review (when applicable) but is not the Version of Record and does not
reflect post-acceptance improvements, or any corrections. The Version of Record is available online at: 
https://doi.org/10.1140/epjc/s10052-024-13207-5

\providecommand{\href}[2]{#2}\begingroup\raggedright\endgroup

\end{document}